\documentclass[useAMS,usenatbib]{mn2e}
\usepackage{epsfig}
\usepackage{graphicx}
\usepackage{psfrag}
\usepackage{amssymb}
\usepackage{amsmath}
\usepackage{color}
\usepackage{url}
\usepackage{enumerate}
\usepackage{times}

\definecolor{light-gray}{gray}{0.7}

\DeclareMathOperator{\diag}{diag}

\usepackage{mathrsfs}

\newcommand{\imshape}{\textsc{im3shape}}
\newcommand{\Imshape}{\textsc{Im3shape}}

\title[]{\imshape: A maximum-likelihood galaxy shear measurement code for cosmic gravitational lensing}
\author[J. Zuntz et al.]
{Joe Zuntz,$^{1234}$\thanks{E-mail: jaz@star.ucl.ac.uk} Tomasz Kacprzak,$^{1}$  Lisa Voigt,$^{1}$
 Michael Hirsch,$^{1}$
Barnaby Rowe$^{1}$ \newauthor and Sarah Bridle$^{12}$\\
$^{1}$Department of Physics \& Astronomy, University College London, Gower Street, London, WC1E 6BT\\
$^{2}$Jodrell Bank Centre for Astrophysics, School of Physics and Astronomy, The University of Manchester, Manchester M13 9PL \\
$^{3}$Astrophysics Group, University of Oxford, Denys Wilkinson Building, Keble Road, Oxford OX1 3RH \\
$^{4}$Oxford Martin School, University of Oxford, Old Indian Institute, 34 Broad Street, Oxford OX1 3BD\\
}

\begin{document}
\maketitle
\begin{abstract}
  We present and describe \imshape{}, a new publicly available galaxy
  shape measurement code for weak gravitational lensing shear.  \imshape{} performs
  a maximum likelihood 
  fit of a bulge-plus-disc galaxy model to
  noisy images, incorporating an applied point spread function.  We detail
  challenges faced and choices made in its design and implementation,
  and then discuss various limitations that affect this and other maximum likelihood
  methods.  
  We assess the bias arising from fitting an incorrect galaxy model using simple noise-free images and find that it should not be a concern for current cosmic shear surveys.
 We test \imshape{} on the GREAT08
  Challenge image simulations, and meet the requirements for upcoming cosmic shear surveys in the case that the simulations are encompassed by the fitted model, using a simple correction for image noise bias.
For the fiducial branch of GREAT08 we obtain a negligible additive shear bias and sub-two percent level multiplicative bias, which is suitable for analysis of current surveys. We fall short of the sub-percent level requirement for upcoming surveys, which we attribute to a combination of noise bias and the mis-match between our galaxy model and the model used in the GREAT08 simulations.
We meet the requirements for current surveys across all branches of GREAT08, except those with small or high noise galaxies, which we would cut from our analysis. 
  Using the GREAT08 metric we 
  we obtain a 
 score of Q=717 for the usable branches, 
 relative to the goal of Q=1000 for future experiments.
 The code is freely available from https://bitbucket.org/joezuntz/im3shape.
\end{abstract}


\section{Introduction}
%
%

Weak gravitational lensing 
has the most potential 
for
constraining the cosmological parameters that describe the dark
universe (Albrecht et al. 2006; Peacock \& Schneider 2006). As dark
energy slows the growth of dark matter structures, these in turn affect
the appearance of distant galaxies, distorting them slightly as
gravitational fields deflect their light.  This distortion is known
as \emph{cosmic shear}. Its measurement from individual galaxy images
is crucial to the field of weak gravitational lensing, as the amount
of distortion is directly related to the amount of intervening dark
matter. However, the stretch induced by cosmic shear is typically on
the order of a few percent, which renders accurate shape measurement
an intricate but also intriguing problem.

%
%

Several upcoming and future observational surveys plan to capitalise
on the potential of cosmic shear for discovering the nature of dark
energy.  These include imminent ground-based projects (the
Kilo-Degree Survey, 
Hyper Suprime-Cam\footnote{\url{http://www.naoj.org/Projects/HSC/HSCProject.html}} and
the Dark Energy Survey:
DES\footnote{\url{http://www.darkenergysurvey.org}}), ground-based
telescopes under construction (the Large Synoptic Survey Telescope:
LSST\footnote{\url{http://www.lsst.org}}), and future space telescopes
(\emph{Euclid}\footnote{\url{http://sci.esa.int/euclid}} and
\emph{WFIRST}\footnote{\url{http://wfirst.gsfc.nasa.gov}}).

It is perhaps surprising that a major challenge for these experiments
is simply measuring the shapes of the galaxy images.  The difficulties
arise because the galaxies involved are: small, less than the Point
Spread Function (PSF) size in many cases; noisy, with a signal-to-noise
ratio (S/N) of $\sim 10$; often irregular; and convolved with a PSF which
itself induces ellipticity in the galaxy that is typically greater than the
gravitational shear we wish to measure.  To fulfill the
potential of cosmic shear these problems must be overcome, allowing
precise and accurate measurements of the underlying gravitational shear.

\subsection{Existing methods}

The best known and most commonly-used method in weak lensing shape
estimation remains that described by \citet{kaiseretal95},
\citet{luppinokaiser97} and \citet{hoekstraetal98}, known as KSB or
KSB+ (see also \citealp{schneider06} for an overview).
KSB uses weighted quadrupole moments on
galaxy images and PSF images to provide an estimator of gravitational
shear.  Many implementations of this approach, and derived methods
adapted for space-based data (e.g.\ \citealp{rhodesrg00},
\citealp{schrabbacketal10}), have been used to estimate gravitational
shear over more than a decade.  An important discovery of the Shear TEsting Programme (STEP)
competitions was the sensitive dependence of the accuracy of KSB upon
the details of its implementation (see Section \ref{sect:contests}, below; \citealp{STEP1mnras,STEP2mnras}).

Recent modifications to KSB have sought to improve accuracy by adding
higher-order terms to series approximations in the method (e.g.\
\citealp{okurafutamase09,violaetal11,melchioretal11,okurafutamase12}).
More fundamentally different shape measurement approaches have also
been proposed.  A relatively early example was the `commutator' method
of \citet{kaiser00}; although this method has not found broad adoption
in practice,
the work contained a number of insights that were to inform the later
development of the field.

Gaussianization, or Re-Gaussianization, approaches (e.g., REGLENS: see
\citealp{shera}, Section 4.3;
\citealp{mandelbaumea2005a,hirataseljak03}) first convolve galaxy
images with an additional kernel designed to make the final PSF as
Gaussian, and isotropic, as possible. Such images then closely match
many of the implicit assumptions in weighted quadrupole moment approaches such as
KSB, so moments can then be used to extract shear estimates with
reduced bias.  Gaussianization methods remain competitive in terms of
systematic bias \citep{mandelbaumea2005a,shera} and have been used
successfully in a number of studies in the Sloan Digital Sky Survey
(e.g.\ \citealp{huffetal11,reyesetal10,Mandelbaum,hirataea04}).

A family of methods known commonly as \emph{shapelets} construct a
model of PSF-convolved galaxy images as a sum of orthonormal
Gauss-Lagurre polynomial functions (e.g.\
\citealp{bernsteinjarvis02,refregierb03,masseyr05,kuijken06,Masseyetal,nakajimab07}).
Effectively
the ellipticity of derived shapelet models can then be used to provide
an estimator of gravitational shear.  \citet{melchior2010} raised some
conceptual concerns with the approach, and while similar (perhaps more
drastic) conceptual problems with KSB did not affect its popularity
there are fewer examples of shapelet shear estimation from real data
in the literature (although see \citealp{velanderetal11}).

The number of proposed shape measurement methods is steadily growing.
Interest in the problem has undoubtedly been stimulated by impending
survey data of unprecedented statistical power, and by
the blind analysis contests we describe in Section
\ref{sect:contests}, below.  
Examples of innovative recent proposals the
competitions described in the next section, and include Fourier Domain Null Testing
\citep{bernstein10} and shear estimation via lookup table (e.g.,
`MegaLUT'; see the review in the appendices of \citealp{great10resultsmnras}).

The final family of shape measurement methods 
are commonly known as model-fitting methods, and involve the
comparison of simple parametric galaxy models (convolved by the PSF)
to imaging data. The galaxy model ellipticity, among other parameters
inferred in the model fit, make up the shear estimator. 

The {\it{lens}}fit
\citep{millerkhhv07,kitchingmhvh08, milleretal12} method seeks to perform Bayesian
inference on galaxy model parameters and performed well in the
GRavitational lEnsing Accuracy Teasting (GREAT08) Challenge: see Section
\ref{sect:contests}, below;
\citep{great08resultsmnras}. The  more recent DeepZot method
that also used a model-fitting algorithm proved the eventual winner of
GREAT10 
\citep{great10resultsmnras}.  

The im2shape public code
\citep{im2shape}\footnote{\url{http://www.sarahbridle.net/im2shape/}}
modelled both the galaxy and the PSF as a sum of Gaussians, inspired by~\citet{kuijken99}.  It
performed MCMC sampling and took the mean of the 
of the ellipticity values of the samples as the shear estimate.
It was used by \citet{bardeauea05,bardeauea07} to measure galaxy cluster masses.

\subsection{Shear measurement contests}\label{sect:contests}

Researchers have come together to assess these many shear measurement
methods in different regimes.  The Shear TEsting Programme (STEP)
was a joint blind analysis of simulated data by groups from within the shear measurement
community \citep{STEP1mnras,STEP2mnras}.  It showed that the shear
measurement problem is far from trivial, but demonstrated that the cosmological
constraints set at that time were not limited by shear measurement
accuracy.  The first two GRavitational lEnsing Accuracy Testing
(GREAT) Challenges posed the problem to computer scientists and
provided sufficient simulations to assess whether methods were
suitable for future surveys~\citep{great08mnras,great10handbookmnras}.
They showed that there are some regimes where existing methods work
well enough, but others where further progress is
necessary~\citep{great08resultsmnras,great10resultsmnras}. The third
challenge in the series,
GREAT3,\footnote{\url{http://great3challenge.info}} is currently in
progress towards a challenge opening and simulation data release in
2013.  In addition to a large simulation dataset, GREAT3 is leading
the collaborative development of an open source extragalactic image
simulation software
toolkit,\footnote{\url{https://github.com/GalSim-developers/GalSim}} 
further advancing the STEP and GREAT
programme of providing the community with a common reference for
comparison and improvement of galaxy shape measurement methods.

\subsection{\Imshape}

This paper presents a successor code to the im2shape code described
above, called \imshape.  It uses (by default) a sum of two Sersic
profiles for the galaxy, similar to
{\it{lens}}fit~\citep{millerkhhv07,kitchingmhvh08,milleretal12},  and
a Moffat profile for the PSF.  
It calculates the
maximum likelihood parameter values and corrects for noise
bias~(see, e.g.,
\citealp{refregierkabr12,melchiorviola12,hirataea04,bernsteinjarvis02}) 
using the calibration scheme described in
\citet{kacprzakzrbravh12}.

In Section \ref{requirements section} we discuss the requirements for
shape measurement for current and future data sets.  Then in Section
\ref{code section} we present the \imshape{} code that we are
releasing, the details of its operation and the choice of parameters
it uses.  Section \ref{noise-free section} is concerned with
noise-free simulations, to verify the code in ideal circumstances and
to explore what biases the use of an incorrect model can introduce to
the results.  We present the results of tests of the code on the
GREAT08 simulation set in Section \ref{sec:great08}, and conclude in
Section \ref{conclusions}.  We give more details of the code
implementation in Appendix A, of the termination criteria in Appendix
B and overview the analysis process in Appendix C.
Appendix~\ref{sec:jacobian} gives the equations and details used for
analytic calculation of likelihood gradients with respect to the
parameters.

\section{Requirements}
\label{requirements section}

Following~\citet{STEP1mnras}, the accuracy of shear measurement
methods are generally 
quantified in the literature in terms of the multiplicative error $m$
and additive error $c$ on the true shear $\gamma^{\rm t}$, such that
\begin{equation}
\hat{\gamma}_i=(1+m_i) \gamma^{\rm t}_i+c_i
\end{equation}
where $\hat{\gamma}_i$ is the shear estimate and the subscript $i$
refers to the two shear components. It is assumed there is no cross
contamination between, for example, $\hat{\gamma_1}$ and 
$\gamma^{\rm t}_2$. 

The requirements on $m$ and $c$ for cosmic shear are computed in~\citet{amarar08}
as a function of the survey area, depth and galaxy number
density~\citep[see also][]{kitchingetal09}. \citet{amarar08} consider
a two-point statistical analysis of the shear field, and the
requirements are set so that the systematic error equals the
statistical error, thus providing an upper limit on the allowed bias.
                                                                                                  
The GREAT08~\citep{great08mnras} challenge quantifies the accuracy of
shear measurement methods using a single number called the
\emph{quality factor}, $Q$. The equation relating $Q$ to $m$ and $c$
is given in \citet{voigt&bridle10}.
A related but different definition is used in GREAT10~\citep{great10resultsmnras} but we stick to the GREAT08 definition here because we use the GREAT08 simulations.

In Table~\ref{tab:req} we show the requirements on $m$, $c$ and $Q$
for three sets of survey parameters, chosen to represent current,
upcoming and future cosmic shear surveys.  Examples of surveys with
requirements comparable to the `current' set of requirements are the
Canada-France-Hawaii Telescope Lensing Survey (CFHTLenS:
\citealp{heymansea12}), and likely the first results from KIDS, DES
and HSC \citep{dejongea12,des05,takada10}.  
The upcoming survey requirements correspond roughly to the
full DES and HSC surveys.  The future survey requirements correspond
to LSST and \emph{Euclid} (e.g.\ \citealp{changea12,laureijsea11}).

\begin{table}
\center
\caption{Shear measurement requirements on current, upcoming and future surveys.
Requirements on the shear multiplicative and additive bias parameters $m_i$ and $c_i$, and the GREAT08 $Q$ value are shown for survey parameters: area $A$, number of galaxies per square arcminute $n_{\rm gal}$ and median redshift $z_{\rm m}$. 
}
\label{tab:req}
\begin{tabular}{|c|ccc|ccc|}
\hline
Survey &  $A$ & $n_{\rm gal}$ &  $z_{\rm m}$ & $m_i$ & $c_i$ & $Q$ \\
\hline
Current & 170 & 12 & 0.8 & 0.02 & 0.001 & 43\\
Upcoming & 5000 & 12 & 0.8 & 0.004 & 0.0006 & 260\\
Future & $2 \times 10^{4} $ &  35 & 0.9 & 0.001 & 0.0003 & 990\\
\hline
\end{tabular}
\end{table}

\section{Im3shape methodology \& code}
\label{code section}
The code we have developed and present in the remainder of this paper
is called \imshape{}.  It belongs to the family of
model-fitting methods: it forward-fits a galaxy model to each
data image it is supplied with and reports the parameters of the
best fitting
model, including the ellipticity components. For finding the
best fit we use a maximum likelihood approach, i.e. we search for
those model parameters that minimize $\chi^2$, the summed square
difference between our generated model and the actual pixel data,
weighted according to a user-supplied weight image. For details on the maxmimum likelihood estimation see Appendix \ref{parameterization-section} and \ref{sec:jacobian}.

Throughout we assume that we have selected a square portion of the image, with {\it stamp\_size} pixels on a side.

\subsection{Code layout}
\imshape{} is a modular C code, with a significant amount of Python glue code
that enables the setting up of new models and their options
automatically.
Key processes like optimization and galaxy model
generation routines are encapsulated in complex structures with simple
interfaces, rather than complicated functions.  This architecture
makes it particularly easy to design new galaxy or star models to be
fit to data.  Since the process of fitting to a large number of
postage stamps is inherently trivial to parallelize, the core of the
code is all single-threaded.

\subsection{Model choice}
Using the flexible systems of \imshape{} we have experimented with
fitting a variety of models, mainly to the images in the GREAT08
challenge.  

We chose by default
a two-component
S\'{e}rsic, with a de Vaucouleurs bulge and an exponential disc, with
the same centroid and ellipticities. In addition we simplify the fitting by fixing the ratio of scale radii, at unity.\footnote{This is similar but not
  identical to the model adopted in the GREAT08 simulations.
  \imshape{} should therefore work reasonably well on those images,
  and we note that this model was originally chosen to be a reasonable description of
  real galaxies.}

  Within that general model we can choose different parameterizations for some components - these are discussed in detail in Appendix \ref{parameterization-section}.

Our complete set of parameters is shown in Table \ref{default parameter table}.  The parameters marked as fixed were kept at default values during our standard analysis but can be varied easily if desired or to study model bias.

\begin{table}
	\begin{center}
			\begin{tabular}{ c | l | c  }	
				Parameter & Meaning & Fixed \\
				\hline
				$x_0$ & Horizontal centroid &  \\
				$y_0$ & Vertical centroid &  \\
				$e_1$ & x-y shear &  \\
				$e_2$ & $45^\circ$ shear &  \\
				$r_d$ & Disc half-light radius &  \\
				$A_b$ & Bulge peak flux &  \\
				$A_d$ & Disc peak flux &  \\
				$R_r$ & Bulge-disc size ratio &  \checkmark \\
				$n_d$ & Disc S\'{e}rsic index &  \checkmark \\
				$n_b$ & Bulge S\'{e}rsic index &  \checkmark \\
				$\Delta e$ & Bulge-disc ellipticity &  \checkmark \\
				$\Delta \theta$ & Bulge-disc angle &  \checkmark \\
		\end{tabular}
		\caption{The complete set of parameters used in the default galaxy model.  Fixed parameters were not by default varied during parameter optimization.}
		\label{default parameter table}
		
	\end{center}
\end{table}

\subsection{Model fitting}
Generating the model image described in the previous Section is not
trivial: we need to ensure that the model is at high enough resolution
to capture sharp central galaxy features.  This is discussed in
Section \ref{upsampling-section}.

Assuming images of the galaxy model can be accurately generated (see
Section \ref{upsampling-section}),
maximizing the computed likelihood (i.e. minimizing the weighted,
squared difference
between the non-linear model and the data) is technically and computationally
challenging.  Reaching the current level of speed and accuracy in
maximization within \imshape{} required considerable
exploration of maximization algorithms and their settings. We therefore give a brief overview of our
findings in Appendix \ref{sec:maximization}, and an overview
of the whole process of image generation and fitting in Appendix \ref{process-section}.

\begin{figure}
\begin{center}
\includegraphics[width=80mm]{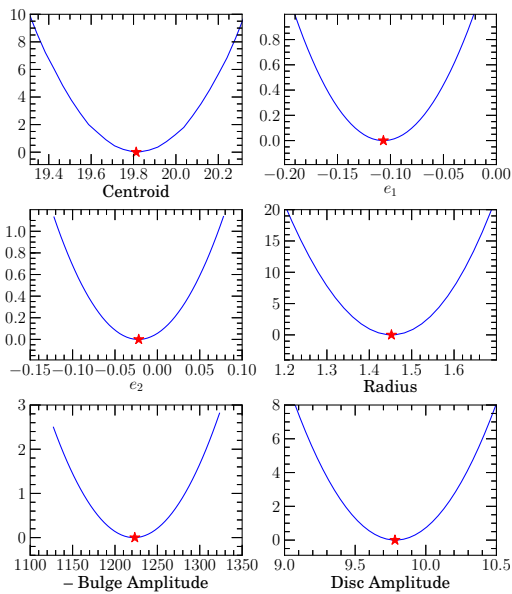}
\caption{Slices in minus log-likelihood in six of the principal parameters (see section \ref{parameterization-section}) around the best fit point found for one galaxy by the likelihood maximizer, which is shown by a star.  
}
\label{like-slices}
\end{center}
\end{figure}

With tuned parameter settings the maximizer, which uses a
Levenberg-Marquardt algorithm, works well.  Figure
\ref{like-slices} shows example likelihoods from the first galaxy of
set 3 in GREAT08. It can be seen that the point is indeed a good
minimum (and further searches can confirm it to be a global one).  The
asymmetry in some of the parameters is part of the cause of the noise
bias phenomenon \citep{kacprzakzrbravh12}.

\section{Noise-free simulations}
\label{noise-free section}
Generating the model image described in the previous Section is not
trivial; for example, we need to ensure that the model is at high enough resolution
to capture sharp central galaxy features without aliasing.  We also
need to understand how best-fitting models may be biased in cases
where they cannot represent the data perfectly.
In this Section we investigate the behaviour and performance of
\imshape{} on noise-free simulations to find optimal parameter
settings for model generation, and analyse in a simple way what potential biases can
arise when the model is insufficiently flexible to describe the data.

\subsection{Upsampling biases}
\label{upsampling-section}

\begin{figure*}
\begin{center}
\includegraphics[width=100mm]{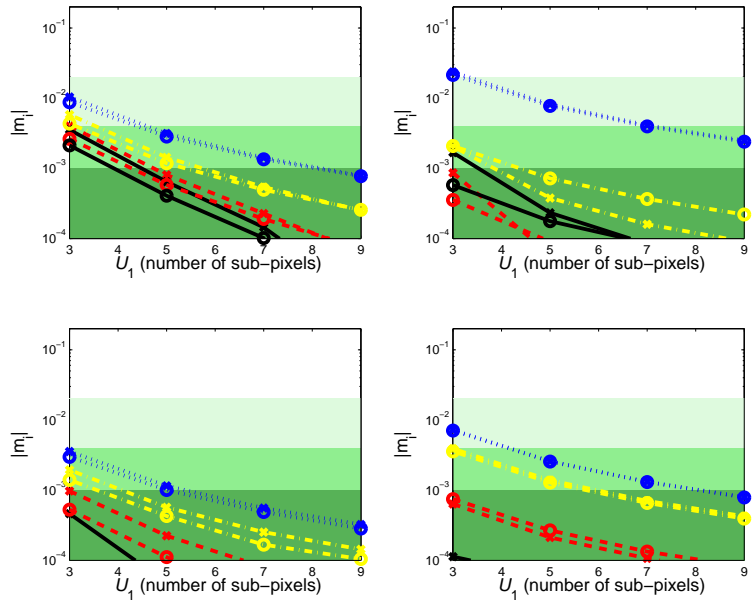}  
\caption{Resolution requirements:  multiplicative bias ($m_1$ crosses,
  $m_2$ open circles) as a function of the number of sub-pixels $U_1$ used
  in the convolution between the galaxy model and the PSF model. Black
  solid, red dashed, yellow dot-dashed and blue dotted lines are for integration of the galaxy flux 
within 7x7, 5x5, 3x3 and 1x1 
sub-pixels around the centre of the galaxy. The flux is integrated over a 7x7 grid within each sub-pixel. 
Top panels: small ($R_{gp}/R_p=1.14$) galaxies. 
Bottom panels: fiducial size ($R_{gp}/R_p=1.4$) galaxies. 
Left panels: de Vaucouleurs galaxies.  
Right panels: exponential galaxies. 
The shaded regions show the requirements on $m_i$ given in
Table~\ref{tab:req}. The upper edge of each shaded region
(from bottom to top) shows the upper limit on the bias allowed for
future, upcoming and current surveys respectively. 
}

\label{fig:upsampling-fig} 
\end{center}
\end{figure*}
The \imshape{} software uses the Discrete Fourier Transform (DFT) to
render images of convolved galaxy profiles: this can be done with
speed thanks to the Fast Fourier Transform (FFT) algorithm, freely
available in highly efficient implementations \citep{FFTW05}.
However, representing the convolution of telescope PSFs with analytic,
continuous, non-bandlimited functions such as the S\'{e}rsic profile
requires careful numerical approximation, as can (if the PSF is also
not bandlimited) the subsequent rendering of convolved profiles into
pixellated images (see, e.g., \citealp{marks09}).  Decisions must be made about how and where to make
approximations in the representation of these profiles, balancing
precision against computational cost.

Upsampling, which is the more accurate rendering of model images using
higher resolution (upsampled) intermediate images for internal
operations such as convolution (see, e.g., \citealp{roweetal12}), is a key tool to eliminate defects in
modelling due to aliasing.
We therefore investigate the requirements on the upsampling parameters
used by the image generation code in \imshape{}.
These three resolution parameters are: 
\begin{enumerate}[(1)]
\setlength{\itemindent}{0mm}
\setlength{\itemsep}{2mm}
\item $U_1$, called {\it upsampling} in parameter files - the
  convolution between the galaxy model and the PSF model is performed
  on a grid of $(U_1\times{\it stamp\_size})^2$ pixels in size,
  i.e. each image pixel is divided into $U_1$ sub-pixels across. The
  flux within each sub-pixel is computed assuming the intensity is
  constant within it and equal to the intensity at the centre of the
  sub-pixel, for all pixels except the central $N_u^2$ pixels (see
  (2)).
\item $N_u$, called {\it n\_pixels\_to\_upsample} in parameter files -
  within a central galaxy region of $N_u^2$ image sub-pixels the flux is
  integrated on a even finer grid within each sub-pixel (instead of
  assuming a constant intensity across each sub-pixel).
\item $U_2$, called {\it $n\_central\_pixel\_upsampling$} in parameter
  files - determines the number of integration intervals within a
  sub-pixel in the central image region; therefore the overall
  resolution in the central galaxy region is $U_1\times U_2$ times the
  data image resolution.
\end{enumerate}
The required values of the these resolution parameters depend on the survey
(see section~\ref{requirements section}). In
figure~\ref{fig:upsampling-fig} we plot the multiplicative bias as a
function of the $U_1$ parameter for different values of $N_u$ and with
$U_2=7$. (The additive bias is negligible in comparison to the
multiplicative bias
because we use a circular PSF for this investigation.) Results are shown for a pure bulge and a pure
disc with $R_{gp}/R_p=1.4$ and 1.14, which are similar to the fiducial
and small size branches used in GREAT08 \citep[see][including for size definition]{great08resultsmnras}.
The bias is computed using a ring-test~\citep{nakajimab07} with 4
angles, $\{0,45,90,135\}$ degrees.

The multiplicative bias is obtained by shearing each galaxy in the
ring by an amount $\gamma_{1}=0.02$ along the $x$-axis and
$\gamma_{2}=0.01$ along the line $y=x$. The PSF is a circular Moffat
profile with FWMH=2.85 pixels and shape parameter $\beta=3.5$. The
postage stamp size (\emph{stamp\_size} in parameter files) is 25
pixels. We find that the requirements on the resolution parameters do
not change if we increase the stamp size to 45 pixels. The bulge-to-disk scale size
ratio, $R_r$, is fixed to 1 during the fitting.

We find that for an upcoming survey (medium green shaded area in
figure~\ref{fig:upsampling-fig}) we require $U_1 \geq $5, $N_u \geq 3$,
and $U_2\geq $7.  For $U_2<7$ the results are unstable.

\subsection{Model bias}
\label{model bias section}
Having established that the resolution parameters described above are
high enough to remove any bias from aliasing effects, we now turn to
the issue of \emph{model} or \emph{external} bias arising from fitting
an incorrect model to the data.  
If a forward model approach could be perfectly implemented and the
posterior distribution accurately characterized for each galaxy (rather than
being reduced to a single point-estimate like the maximum likelihood) and
combined appropriately, then this would be the only remaining source
of bias for simple images like those in GREAT08.  In practice, 
there may be other problems such as neighboring galaxies and image artefacts.

Very realistic galaxy models are almost certainly out of reach of
forward methods for the forseeable future - modelling spiral arms, for
example, is hard enough for a single high-resolution image, let alone
$10^8$ noisy ones.  We defer testing on extremely realistic models
(for example using the {\sc SHERA} simulation code of \citealp{shera}, or
Galsim), and instead apply a necessary but not sufficient test -
fitting a simple but incorrect model using \imshape.

Again, we perform these tests without noise, to separate the issue of
\emph{noise} bias on maximum-likelihood estimates from \emph{external}
bias; of course, there may be an interaction between the two.  As we
only touch upon these issues here, we refer the interested reader to
\cite{voigt&bridle10} and \cite{melchior2010} for a more extensive
disucssion.

Except where otherwise noted, our galaxy simulations in this section used $R_{gp}/R_p=1.4$, equal component scale radii, and half the flux in each of the bulge and disc components $F_B/(F_B+F_D)=0.5$.  The PSF is a Moffat with $e_1=e_2=0.03$, $\beta=3$, and FWHM$=2.85$.

\subsubsection{S\'{e}rsic Indices}
We first test a simple difference between the input (true) model
and the fitted one - a different index of the S\'{e}rsic profile.  Our
input (true) model is generated with a single component, a S\'{e}rsic
profile with index $n_s$, however, we fit our standard bulge plus disc
model.

\begin{figure*}
\begin{center}
\includegraphics[width=120mm]{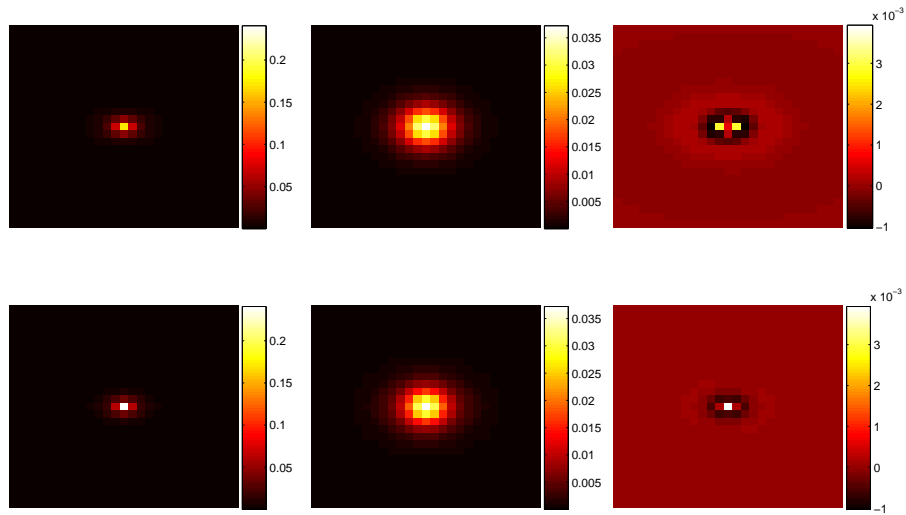}
\caption{ Examples of model bias: \imshape{} co-elliptical bulge
  ($n_s=4.0$) plus disc ($n_s=1.0$) model fits to simulated
  single-component S\'{e}rsic galaxies with $n_s=1.7$ (top panels) and
  $n_s=3.5$ (bottom panels) . Postage stamp images are shown for the
  simulated galaxy (left-hand panels), the PSF-convolved simulated
  galaxy (middle panels), and the residuals between the simulated
  galaxy and the best-fit galaxy model from \imshape{} (right-hand
  panels).  The simulated galaxy has ellipticity components $e_1=0.2$
  and $e_2=0.0$ and the radius along the major axis is 1.84
  pixels. The PSF is a circular Moffat profile with $\beta=3.0$ and
  FWHM=2.85 pixels.
}
\label{summary-fig}
\end{center}
\end{figure*}

Figure \ref{summary-fig} demonstrates this bias: it shows the pre- and post-PSF convolved
simulated galaxy images, and the residuals between the simulated
galaxy image and best-fit galaxy model, for two different values of
the true S\'{e}rsic index. The images are noise-free and the correct PSF
model has been used in the fits.  The error on the true galaxy
ellipticity is approximately $10^{-3}$ for $n_s=1.7$, and
$2\times10^{-4}$ for $n_s\sim3.5$.

The residual image shows different behaviour along the $x$ and $y$ directions, which is what causes the ellipticity to be mis-estimated. As discussed in \cite{voigt&bridle10}, the galaxy size is worst estimated along the direction where the galaxy is smallest relative to the PSF. Therefore the ellipticity is incorrect.

\begin{figure}
\begin{center}
\includegraphics[width=90mm]{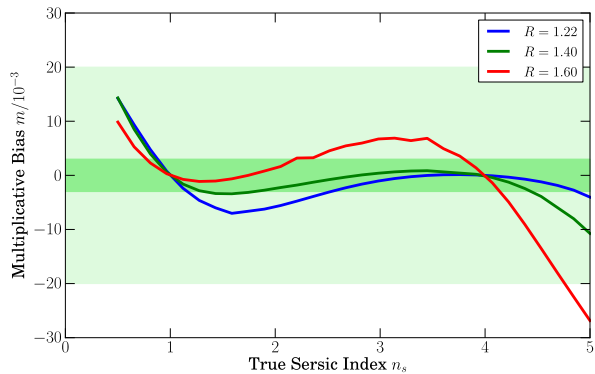}
\caption{The mean multiplicative bias $m$ as a function of true model
  S\'{e}rsic index when fit with a bulge+disc model, for three
  different image sizes as shown.  The shaded bands are the current
  and upcoming survey target biases.  Since the simulations were
  noise-free error bars are negligible.}
\label{m-sersic}
\end{center}
\end{figure}

Our model used in \imshape{} is correct when $n_s=1$ or $n_s=4$, since
at these points the true model is a subset of the \imshape{} model.  We generate
simulations with our fiducial sets of galaxy parameters, except for
the S\'{e}rsic index, which we vary.  We apply a ring test to the
simulations to assess the multiplicative $m$ and additive $c$ biases;
the former is shown in figure~\ref{m-sersic}, a figure for the latter is
omitted as it is negligible in the noise-free case.  As expected, the
bias becomes zero when the true and fitted model coincide at $n_s=1$
and $n_s=4$.

Figure~\ref{m-sersic} suggests that this kind of model bias has the
potential to be a problem for upcoming surveys but is only marginally
relevant for current generation of experiments.  Allowing for
additional flexibility of the model could resolve this potential
issue.  Further tests are needed on more realistic galaxy images to assess the true impact on upcoming surveys.

\subsubsection{Different bulge and disc ellipticity}

\begin{figure}
\begin{center}
\includegraphics[width=90mm]{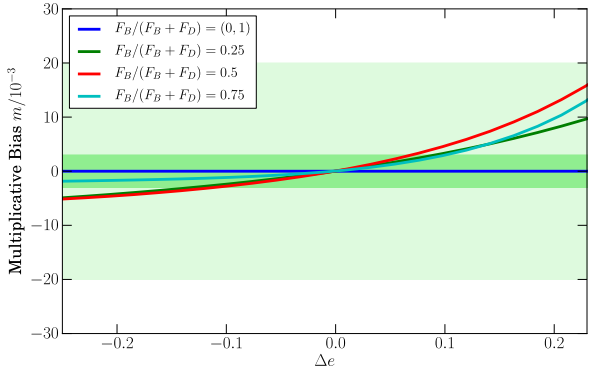}
\caption{The mean multiplicative bias $m$ as a function of true model
  $\Delta e$, the difference between bulge and disc ellipticities,
  when fit with a co-elliptical bulge+disc model.  The shaded bands are target
  biases for current and upcoming experiments.  Since the simulations
  were noise-free error bars are negligible.  In real galaxies $\Delta
  e$ is usually negative.}
\label{m-delta}
\end{center}
\end{figure}

In figure~\ref{m-delta} the multiplicative bias arising from a slightly
more realistic model difference is shown - a bulge and disc model with
components that are not co-elliptical.  This is of interest because it is a difference between the \imshape{} model and the GREAT08 simulations. The input image bulge
ellipticity\footnote{We define ellipticity as $e \equiv
  \sqrt{e_1^2+e_2^2}$ with $e_{1/2}$ from 
  Table~\ref{default parameter table} and Appendix~\ref{parameterization-section}.} $e_b$
is fixed at 0.3 and the disc ellipticity varied so that $e_b =
e_d+\Delta e$. 
The difference between the bulge and disc ellipticity, $\Delta e$, is
varied between -0.25 and 0.25. Typically, bulges are rounder than
discs, i.e $\Delta e < 0$.  The model fit to the data is our standard two-component co-elliptical
profile.

Once again a ring test is applied to assess the bias, and the noise is
zero.  The other galaxy parameters are set to the fiducial values
described above, except that the different curves show different
relative amplitudes of the bulge and disc components.  The bias is
zero when either $\Delta e$ is small or the galaxy is completely
bulge- or disc-dominated, since in that case the secondary component
does not affect the fit at all.  The effect is greatest when the
components have comparable total fluxes, and this bias too 
is small
for the current
generation of experiments, 
if bulges are rounder than discs.

\subsubsection{Incorrect Radius Ratio}

As depicted in Table~\ref{default parameter table}, the ratio between
the half-light radii of the bulge and the disc is fixed by default in
the code.  This is partly to reduce the number of parameters (since
fewer parameters generally means easier optimization), but also
because this particular parameter causes problems with convergence in some regions of
its range. 

This is clearly a potential source of model bias - the ratio of these
radii varies in real galaxies (when bulges and discs can be clearly
identified).  Furthermore, in GREAT08 this value is around $R_r\sim 1.5$. We again apply a ring test to noise-free simulations to
check how severe this problem is.  If we alter the ratio without
changing the disc radius, then the bulge radius changes accordingly.
To separate the effect of fitting small galaxies from this model bias
we keep the ratio of the convolved image to PSF size,
i.e. $R_{gp}/R_p$, fixed at three different values corresponding to
GREAT08 sizes.

\begin{figure}
\begin{center}
	\includegraphics[width=90mm]{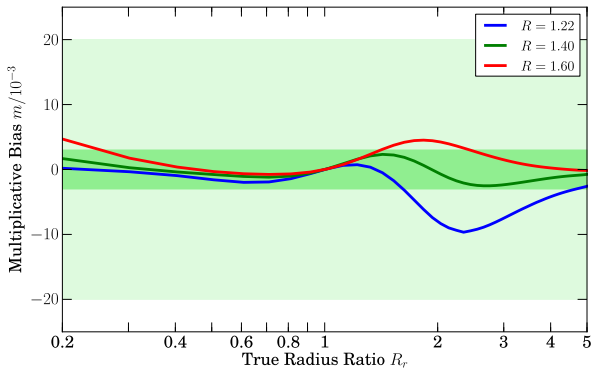}
        \caption{The mean multiplicative bias $m$ as a function of the
          true ratio between the bulge and disc half-light radii for
          various galaxy sizes is shown. Note that the bias arises as
          the ratio has been fixed to one during the fitting with
          \imshape{}. Again, the simulations are noise-free so there
          are no error bars.  The shaded bands are target biases for
          current and upcoming experiments.}
\label{fig:radius-ratio-bias}
\end{center}
\end{figure}

Figure \ref{fig:radius-ratio-bias} shows the results of this test.
If real galaxies have smaller bulge scale radii than discs then the biases are mostly smaller than the requirements for upcoming surveys. If they are larger then the bias would be acceptable for current surveys but problematic for upcoming surveys, with a sign which depends on the galaxy size. For the GREAT08 fiducial branch ($R=1.4$, true $R_r\sim1.5$) we expect a slight over-estimate of the shear. This will combine with the slight underestimate expected from figure 5 for bulges which are rounder than discs, and may partially cancel out to bring results down to be compatible with the requirements for upcoming surveys. We recommend a program of testing on realistic images for calibration of upcoming surveys. 

\section{GREAT08}
\label{sec:great08}

We tested \imshape{} on the Gravitational LEnsing Accuracy Testing
2008 (GREAT08) Challenge ~\citep{great08mnras,great08resultsmnras}
which has a sufficient number of galaxies to test methods at the
accuracy required for future surveys (unlike STEP1 and
STEP2). Furthermore, it features a constant shear across the images
which greatly facilitates the computation of success metrics (unlike
GREAT10) and therfore performance evaluation. The GREAT3 Challenge is
still under construction (Mandelbaum, Rowe et al. in prep).

The GREAT08 Challenge includes two blind competitions: one with low
noise, which has a signal-to-noise ratio\footnote{Here, we adopted the
  GREAT08 definition of signal-to-noise ratio defined by (A11) in
  \citet{great08resultsmnras}, Appendix A, page 14.} of 200
(`LowNoise\_Blind') and one with more realistic signal-to-noise ratio
around 20 (`RealNoise\_Blind'). In this section we show the results of
\imshape{} on these simulations.

\subsection{Low Noise Blind}
The LowNoise\_Blind challenge consists of 15 image files, each of
which has 10,000 galaxy postage stamps of 39$\times$39 pixels. 
All the galaxies are a
mixture of bulge and disk components which are co-centered but not
co-aligned. The misalignment angle is drawn from a Gaussian with a
standard deviation of 20 degrees.  We therefore expect a model bias to
apply in these cases as described in section \ref{model bias section}.
The signal-to-noise ratio of 200 is large enough that noise bias
described in \cite{kacprzakzrbravh12} is expected to be negligible.
The PSF and galaxy size parameters are described in detail
in~\cite{great08resultsmnras}.  In summary, there are 5 image files of
each of 3 different galaxy sizes. This allows us to show how the
success metrics depend on galaxy size.

The GREAT08 images are simulated at essentially infinite resolution by using photon shooting for the PSF and galaxy profiles \citep{great08resultsmnras}. \imshape{} makes its models at high resolution (see section \ref{upsampling-section}).  For this test we ran at two different values of this resolution, with an upsampling 
$U_1=5$ for a coarse run and $U_1=7$ for an improved, higher resolution run that took about twice as long to complete.  We are interested to see both results on the small LowNoise\_Blind dataset and would like to use the coarser resolution for the much larger RealNoise\_Blind dataset.

\begin{figure*}
\begin{center}
	\includegraphics[width=50mm]{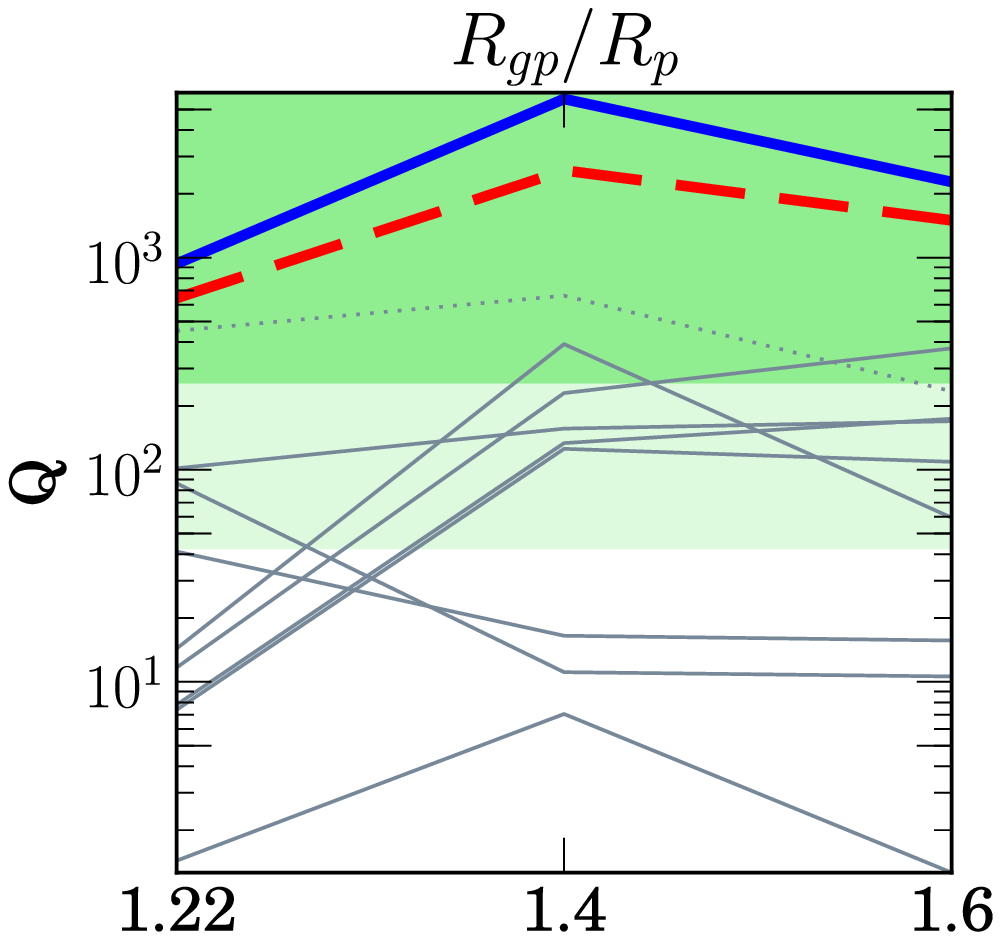}
	\includegraphics[width=50mm]{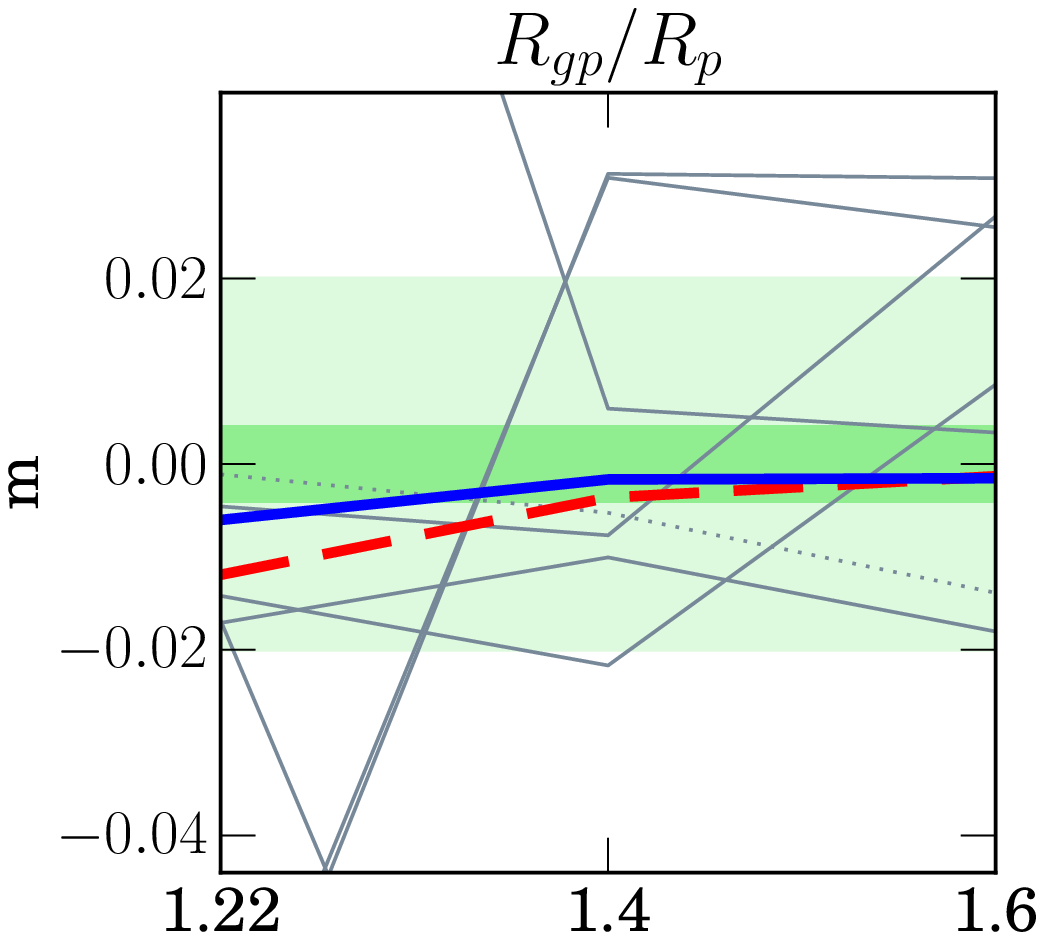}
	\includegraphics[width=50mm]{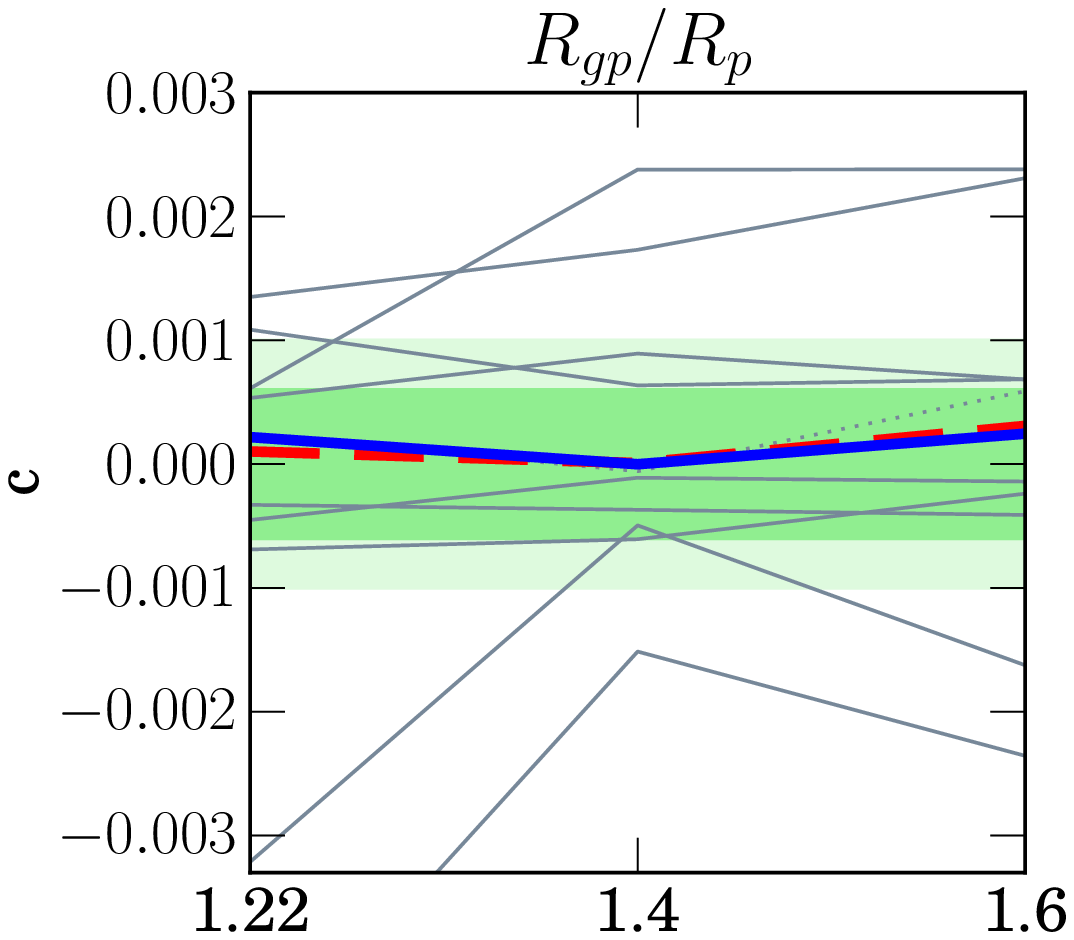}
\caption{The Quality factor score $Q$, multiplicative bias $m$, and additive bias $c$ (in the PSF direction) obtained in the LowNoise-Blind section of the GREAT08 challenge.  The solid blue line is our main result, and the red dashed line is run at a lower resolution.
The green bands are the requirements for current and upcoming surveys. The thin grey lines are other entrants at the time of the challenge, with dots indicating stacking methods.}

\label{fig:lownoise}
\end{center}
\end{figure*}

In figure~\ref{fig:lownoise} we show the \imshape{} results on the LowNoise\_Blind GREAT08 data. At high resolution the resulting $Q$-factor scores reach level $Q>1000$ for all branches, a level at which the statistical noise in the challenge dominates - scores above this level are consistent with zero error.  We reach this score in spite of a model bias: the the GREAT08 simulated galaxies have misaligned isophotes, whereas we fit co-elliptical ones, and the ratio of bulge to disc size ratio is different. 

At the coarser resolution we are within the requirements for future surveys for all but the multiplicative bias on small galaxies, which is nonetheless acceptable for the current generation. We therefore use the coarser resolution in the remainder of this paper.

We also show the results submitted by others at the time of GREAT08. Significant developments have been made in other codes in the meantime, but we can conclude that at that time there were no methods as effective as \imshape{} in this high signal-to-noise regime. The closest competitor is the stacking method by AL, which has not yet been developed into a method that can be used on realistic data. We should however note that the model in the \imshape{} code is very well matched to the GREAT08 simulations, in that they both use exponential disc plus de Vaucouleurs bulge galaxies. An investigation into this topic is beyond the scope of this work, and will be addressed by future challenges including GREAT3.

In running this analysis we determined which parameters in the code can be tweaked to minimize runtime while retaining high accuracy.  The most important such parameters are those controlling the details of the LevMar optimization and convergence behavior.  In subsequent work we have found that tuning these parameters is quite specific to the noise levels and other parameters of the galaxies in question, and we have not been able to draw any general conclusions about them.

\subsection{Real Noise Blind}
\label{sec:realnoise}

The GREAT08 RealNoise\_Blind Challenge is considerably larger, with 2700 sets of $100\times 100$ galaxy stamps.  
The 2700 sets are divided into 9 branches which span a range of simple galaxy models and PSF ellipticities and a range of realistic galaxy sizes and noise levels.
The fiducial branch has a Signal-to-Noise a factor ten lower than in LowNoise\_Blind but the same galaxy size as the central branch of LowNoise\_Blind and the same PSF and galaxy type (co-centred but not co-elliptical bulge plus disc).
Two branches explore the same two extrema in galaxy size as LowNoise\_Blind; two branches vary the Signal-to-Noise by a factor either side of the fiducial value; two branches vary the PSF ellipticity by a factor of two and in direction; and two branches explore variants on the galaxy model: bulge or disc and bulge plus disc with an offset in the center.
For more details on the challenge, see \citet{great08}.

We tuned the maximization parameters described in Appendix \ref{termination-criteria-section} to improve the mean likelihood of all the galaxies in a single set; on the basis that if a set of minimisation parameters did not find the best likelihood then it must not have converged fully. We note that the mean ellipticity obtained for this single set was quite sensitive to the maximization parameters, and therefore we recommend users of the \imshape{} code should perform a similar tuning on new galaxy samples.

\subsection{Noise Bias Calibration}

At the noise levels in the RealNoise\_Blind Challenge we do expect that we need to correct for the noise bias effect described in \cite{refregierkabr12}, \cite{kacprzakzrbravh12} and references therein.  We describe our approach to noise bias calibration in this subsection.

In \cite{kacprzakzrbravh12}  we ran \imshape{} on noisy simulations with a range of input parameters to find the behaviour of the noise bias with galaxy properties.  We then attempted to remove the bias by matching the measured galaxy properties to the simulation input parameters to predict the expected bias, which we subtracted off.    We ran this method on branch 3 of our GREAT08 analysis, which has the same galaxy model as \imshape{}, and found $Q=166$, with $m=3.0\times 10^{-2}$ and $c=7\times 10^{-4}$ which is nearly satisfactory for current experiments but well outside our requirements for upcoming experiments.

This straightforward correction fails to do better because the
observed properties are themselves noisy (and thus noise-biased), and so our bias estimate is itself biased.  In practice, we have found it ineffective to fully calibrate galaxy ellipticity measurements one-by-one using their individual properties, and turn instead to a population-based approach in which a complete class of galaxies is calibrated collectively.

The approach that we take is discussed in Appendix \ref{calibration appendix}.  Briefly, we use the deep data from the low-noise part of the challenge to provide distributions of the galaxy properties, and then perform simulations to compute the mean multiplicative and additive biases for those properties.  We then apply these mean biases to 
the shear estimates from each set.  

To predict biases for a given parameter set we simulate galaxies at
grid points with scale radii $r_d=\{1, 2.5, 5\}$ and bulge flux fractions
$F=\{0, 0.5, 1\}$, and with ellipticity from $e_1=-0.8 $ to $0.8$.  At
each grid point we fit a cubic polynomial to the bias $\hat{e} -
e_\mathrm{true}$ as a function of $e_\mathrm{true}$, and store the
coefficients.  To get the bias for intermediate parameters we use a
Gaussian radial basis function interpolator.  Since the simulations
are costly we simulate only at $S/N=20$, and extrapolate to other
values using the result from \cite{refregierkabr12} and \cite{kacprzakzrbravh12} that the bias
scales
as the inverse of the square of the signal-to-noise.

There are several factors about the structure of GREAT08 that are unrealistic in this context.  Firstly, not every branch in the main sample has a corresponding low-noise version, whereas in real surveys it should be possible to match populations with more care in most cases.  This would tend to make noise calibration harder on GREAT08.  Secondly, the sizes of galaxies in GREAT08 are not drawn from a population - they have a single true value.  This would tend to make calibration more effective than is realistic.  We also note that the information needed to fully perform the calibration in this way was not public at the time of the challenge.

Where possible we match branches using the corresponding low-noise branch; where none exists we use the nearest approximation.  
See Appendix \ref{calibration appendix} for  more details.

\subsection{Scores}

\begin{figure*}
\begin{center}
	\includegraphics[width=160mm]{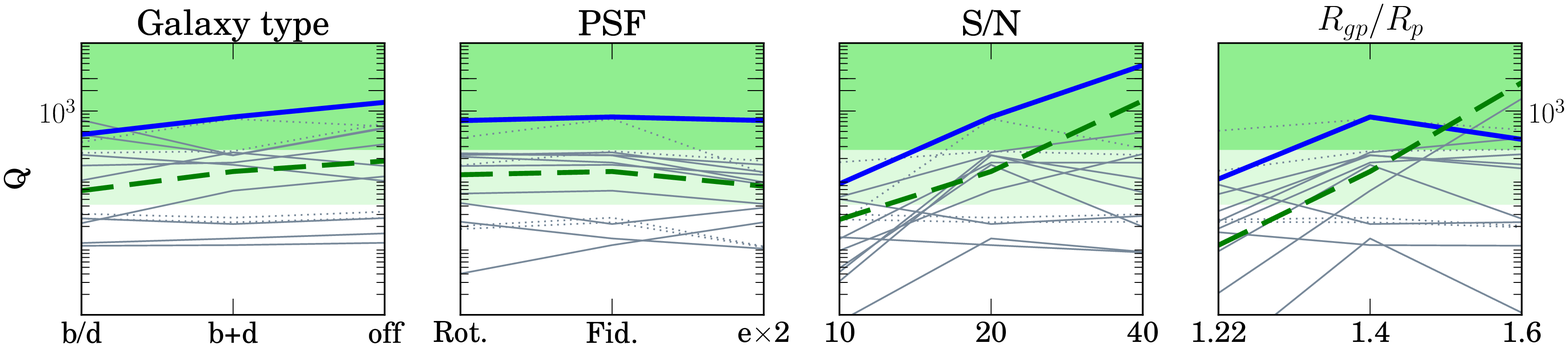}
	\includegraphics[width=160mm]{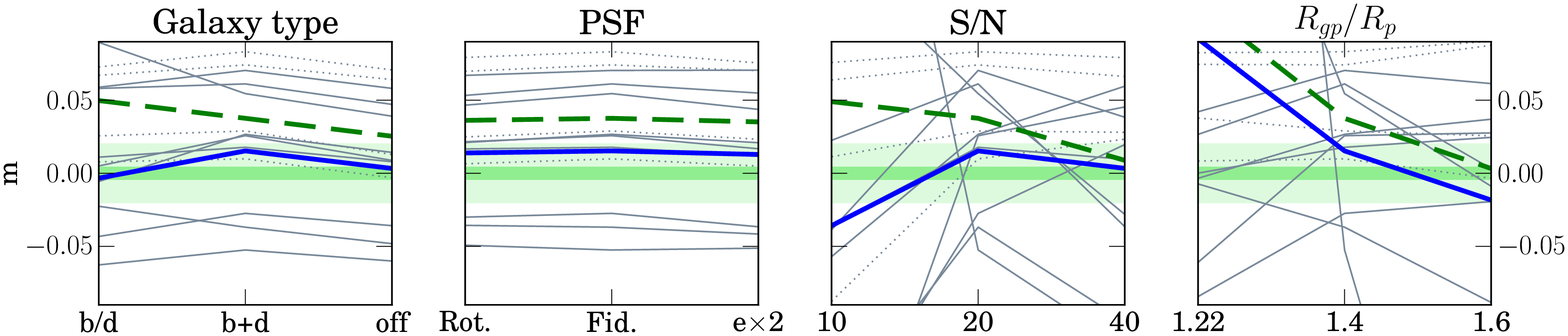}
	\includegraphics[width=160mm]{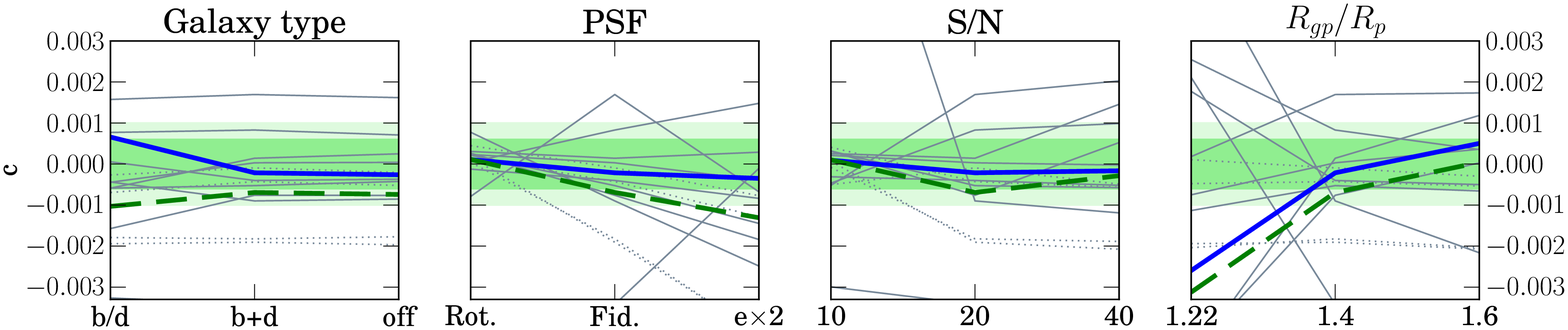}

\caption{The Quality factor score $Q$, multiplicative bias $m$, and additive bias $c$ obtained in the RealNoise-Blind section of the Great08 challenge. Each column shows scores with a different variation from the fiducial galaxy parameters.  The thick solid blue line is our primary result, the \imshape{} scores following the matched-population noise bias calibration described in the text.  The dashed green line is the score without any calibration.  The green bands are the requirements for current and upcoming surveys. The thin grey lines are other entrants at the time of the challenge, with dots indicating stacking methods.  The galaxy types are bulge \emph{or} disc (50\% of each; labelled $b/d$), co-elliptical bulge plus disc ($b+d$) and bulge plus disc with offset centroids (off).  The PSF values were fiducial (fid), rotated by $90^\circ$, and double ellipticity($e\times 2$). }
\label{fig:realnoise}
\end{center}
\end{figure*}

Q-factor scores and $m$ and $c$ values for the pre- and post-noise bias calibration results are shown in figure \ref{fig:realnoise}.  Even before the calibration the bias on the $S/N=40$ galaxies was small enough that we achieved a quality factor $Q>1000$, with $m$ and $c$ small enough for upcoming surveys.  At lower signal to noise, as we expect, the pre-calibration Q factor drops to below 100.

After the noise bias calibration procedure described above the Q factors and the $c$ values reach the levels required for upcoming experiments for most of the branches of the challenge.  The results are stable with PSF type.   The m values are somewhat more variable but, depending on the 
exact branch, can reach the required levels.  In particular the value is good in the b/d branch, where where we expect no model bias.

Our most encouraging result is the stability across the galaxy type branches. The only branch for which we use the correct model is the first galaxy type, where the simulation model is single-component bulge or disc.  For all other branches we expect some model somewhat like the ones discussed in section \ref{model bias section}, but as in that section the effect is not critical for current 
surveys.

The branches for small and noisy galaxies are clearly more problematic, 
and we would be forced to remove such galaxies from any current analysis.
We believe the extrapolation fails badly in the high-noise case because of a slowdown in the magnitude of the bias with ellipticity, which does not follow the $b\sim(S/N)^{-2}$ relation that we used.  Very noisy galaxies have much broader scatter in measured ellipticity, and when these values start to push up to the edge of the space, at $|e|=1$, this acts as an $m<0$ bias which partially counteracts the usual $m>0$ one.  Missing this effect can lead to extrapolated values over-correcting the bias.

\section{Conclusions}
\label{conclusions}

We have presented a public galaxy shape fitting code, \imshape{}, for
weak lensing shear measurement.  Our code efficiently finds the
maximum likelihood fit of our model to any given galaxy, and does so
extremely well: we find little sign of any bias due to poor minimization for simple simulated galaxies, following careful tuning of the model image 
resolution settings 
and other parameters.  The code is modular and easily extensible, and we would welcome community contribution to any part of it\footnote{\url{http://bitbucket.org/joezuntz/im3shape}}.  The technical problems in generating models and fitting them to galaxies are extensive, with particular problems of resolution, parameterization, and 
non-linear numerical optimization.  We discuss our solutions to these problems in the appendices to this paper.			

We find that we need to generate model images at five times the data resolution to reach our desired accuracy for upcoming surveys.  We also find that to accurately capture small scale features in de Vaucouleurs cores we must sample a small central region even more finely - at $35$ times the data resolution.

Our code fits a simple bulge plus disc galaxy model to galaxy images, and we have made an exploration of model bias, which can arise when true galaxies fail to be as neatly analytic in form as our model ones.   We find that reasonable deviations between the 
simple
true and assumed models cause only minor bias, not significant for current data and early results from the upcoming generation of surveys. The bias will, however, attain greater significance as 
upcoming
surveys reach their full statistical potential.  Our tests were rather limited in that we considered only simple differences from our canonical model - deviations in Sersic index, and differences between the size and ellipticity of the two model components.   We look forward to performing more realistic tests with the GalSim tool set in future work (Mandelbaum, Rowe et al. in prep).

We have demonstrated the utility of our method and code by testing on the GRaviational lEnsing Accuracy Testing 2008 (GREAT08) Challenge in both low and realistic noise regimes. 
We find extremely good results for the low noise simulations, with 
multiplicative and additive shear measurement biases 
well within the requirements for upcoming (Stage III) surveys, for the medium sized and large galaxies. For practical reasons we chose to compromise on the resolution we used in the fits and thus our results for the smallest galaxies are only sufficiently good for current surveys. 

In~\cite{kacprzakzrbravh12} we calculated the degradation of shear measurements due to noise on images and discussed correction schemes. We do not expect this to be significant at the signal-to-noise of the LowNoise simulations ($S/N=200$). We expect the main difficulty to arise from resolution and the differences between our fitted model and the GREAT08 galaxy models. Both our galaxy models and the GREAT08 low noise simulations assume galaxies are made of a de Vaucouleurs bulge plus an exponential disc. The difference is that we assume these two components are co-centric and co-elliptical with the same size ratio, whereas in GREAT08 low noise the two components have different ellipticities, are not co-aligned and have a bulge scale radius which is approximately 50\% larger than the bulge scale radius. We find that there is little resulting problem arising from model bias alone, as expected from our simpler investigations. 

We ran \imshape{} on the 27 million galaxies in the realistic noise part of the GREAT08 Challenge, which we chose over GREAT10 because it is simpler and the variable shear fields used in the latter were not relevant to this method. 

At this signal-to-noise ($S/N\sim20$) we do expect significant noise bias due to pixel noise in the images~\citep{kacprzakzrbravh12}, leading to a multiplicative shear measurement error of a few per cent. We found that the correction scheme tested in~\citep{kacprzakzrbravh12} does not work well on GREAT08 due to the bias-on-bias problem described in that work.  We therefore deviated from the methods discussed there, which corrected each galaxy individually, in favour of a global correction using the knowledge of galaxy properties measured in the corresponding low noise sample. The leading order correction to shear power spectrum measurements is expected to simply depend on the mean multiplicative bias of a population, which can be corrected globally. Since the noise bias depends on galaxy properties like size, and galaxies with similar properties can be clustered, then it is possible this scheme may not be satisfactory for high precision corrections.

We scale the noise bias correction from the fiducial value according
to signal-to-noise, as expected in the analytic calculations
of~\citet{refregierkabr12}. This works well for the lower noise images
but less well for those at higher noise levels as also noted
in~\citep{kacprzakzrbravh12}. As stated in that work we are not
satisfied with the performance of our optimiser in this highest noise
regime and would cut such images from cosmic shear analyses at this
point, or calibrate using a more extensive suite of population matched
simulations.  A similar concern applies to the smallest galaxies.

For all other branches of GREAT08 we attain additive and multiplicative biases which are within the requirements for current surveys, and have quality factor metrics which are as good or better than other non-stacking methods at the time of the GREAT08 Challenge.

Further work appears to be required before \imshape{} can be applied to the final years of upcoming Stage III cosmic shear surveys. We speculate that our residual shear measurement bias is due to the combined effects of using the incorrect galaxy model in the fit, and the noise on the images. Ideally an improved shear measurement method could be found that did not suffer from these types of bias. Alternatively the calibration scheme needs to be tailored more accuractly to the analysis in question, for example by using a suite of low noise images with known shears. 

To be fully applicable to real data there are a number of extensions required to the code, for example (i) the ability to handle multiple exposures of the same patch of sky (ii) the capacity to deal with or flag neighboring objects contaminating the shear measurement. 

Furthermore, we have not investigated here the effect of postage stamp size or the effect of a weight map cutting out parts of the galaxy image.
These developments are ongoing but testing them is beyond the scope of the present work.

\section*{Acknowledgments}
We thank Steve Gull, Jean-Paul Kneib, Eduardo Cypriano, Phil Marshall, John Bridle, Patrick Hudelot, Sebastien Bardeau, David Hogg, Alexandre Refregier, Adam Amara, Filipe Abdalla, Niall Maccrann, Emmanuel Bertin, Jim Bosch, Mike Jarvis, Fabrizio Sidoli, Gary Bernstein, Catherine Heymans and Tom Kitching for helpful discussions, and Dugan Witherick for patient and extensive computational help.

JZ, TK, SB, BR, LV and MH acknowledge support from European Research Council in the form of a Starting Grant with number 240672.

\bsp

\bibliographystyle{mn2e}
\bibliography{bib_im3shape}

\appendix

\section{Modelling and parameterization}
\label{parameterization-section}

\subsection{Choice of ellipticity parameters}

The performance of all the optimizers we have tried is heavily dependent on the particular choice of parameterization.  In particular, the choice of how to parameterize the shear components is particularly important since they are our target.  
We define ellipticity to match the effect of shear on images: $e = (a-b)/(a+b)$.
The two obvious parameterizations based on this, $(e_1, e_2)$ and $(e,\theta)$, were both found to be problematic.  

Using the two separate ellipticities $e_1$ and $e_2$ as parameters creates a non-linear boundary at $e_1^2 +e_2^2=1$, the edge of the unit circle.

While most optimization methods cleanly handle linear boundaries (for example simple limits on parameters), non-linear ones are less well served.  We tried simply assigning an extremely low likelihood to any proposed sample beyond this boundary, but fast optimization methods use likelihood differences to determine step sizes, so they typically behave wildly if large negative values are used to signal edges.  Since the images of model galaxies diverge as $|e|\rightarrow1$ simply making the models of these galaxies is also prone to numerical problems.

Using the polar coordinates $(e,\theta)$ as parameters works well at high ellipticity where we use a simple parameter limit on $e$, but fails at lower ellipticities.  As $e\rightarrow 0$ the likelihood becomes a plateau in $\theta$ since the coordinates are degenerate.  Optimizers spent an extremely long time wandering around the circle.

We 
found a reliable parameter set 
which
is a variant of the first parameterisation, $(e_1, e_2)$.  We use  $(e_1, e_2)$ as the parameters varied by the optimizer, but map them to a 
disc wherever they stray beyond it.  We mirror the parameter space radially about some $e$ just below unity (we found 0.95 worked well), so that as the input $e$ increases above the maximum, the 
value given to the minimiser
first decreases and eventually reaches zero, and turns negative so that the signs of the output $e_1$ and $e_2$ are flipped.  Although this mirror structure causes repetition in the likelihood surface and multiple modes we have not found this to be an issue in practice.

\subsubsection{Other parameters}

There are various radius parameters we could choose for each of the components.  The choice is particularly important for the sharp bulge profiles, as their full-width half-maxima are extremely small and so they are not suitable parameters.

A better choice that works for both components is the radius which encloses half the total flux of the complete profile. Rather than using two radius parameters, we the disc radius and the ratio between the radius of the bulge and disc.  We find (see section \ref{model bias section}) that no large model bias is generated by fixing the latter parameter provided it is approximately correct.  We therefore use a fixed value for any given data set.  For our GREAT08 runs this value was arbitrarily set to unity, to avoid trying to match too closely to the GREAT08 truth values.

A likelihood plateau occurs for small objects like stars, where the shears are irrelevant.  In these regimes the fitted radii of the objects tended to fail to converge as they varied around zero.  We solved this problem by setting a small transition radius $R=0.4$ sub-pixels and a replacing the radius $r$ with $\frac{1}{2} (R+r^2/R)$ when $|r|<R$ and $|r|$ otherwise.  The minimum value radius actually used is then $R/2=0.2$ sub-pixels.

\subsubsection{Marginalization}

We have experimented with marginalization over the amplitude components of the models.  Since the model is linear in these parameters their likelihood through a given slice in the other parameters is Gaussian and we can analytically marginalize over them.  We have found this to be slightly problematic in general - the marginalization is not particularly numerically stable and since we have two components we need to perform the PSF convolution twice if we sum them after marginalization.  These two effects meant it was in fact slower to use this trick, provided that we have a reasonable initial estimate of the ellipticity, particularly as our gradient-based minimizer finds the linear parameters very simple to optimize.

If we move the modelling into Fourier space we could further marginalize over the centroid parameters $x_0$ and $y_0$ of the model - combining these dimensionality reductions could well yield a significant speed up.  The downside of this is decreased flexibility in the modelling.  We defer further discussion of this issue.

\subsubsection{Point spread function (PSF)}
\imshape{} has the general facility to use an arbitrary image for PSF, but by default we read a catalog of either Moffat or Airy function parameters and use that to build a PSF image.  It is built at the same upsampled resolution as the image itself, and transformed and stored in Fourier space for faster convolution with trial images.

When these images are made the convolution is done using the FFTW package \cite{FFTW05}.

\subsubsection{Fourier Sampling Kernel}
\label{sec:top-hat}
Our model for the pixels in an image is really a constant value across the surface of the pixel. 
But as far as a Discrete Fourier transform is concerned a 2D array represents a grid of $\delta-$function samples from the image. 
To account for this difference our Fourier-space model of the image needs to include convolution with a square top-hat function the size of a pixel - this is true even when simulating at the same resolution as the data. Since we are also convolving with a PSF we can fold together these convolutions into a single Fourier-space product.  
The Fourier transform of a 2D top-hat is given by:

\begin{equation}
F_{ij} = sinc \left( \frac{i \pi u}{2 n_x}\right)  sinc \left(\frac{j \pi u}{2 n_y}\right) , 
\end{equation}
where the image dimensions are $(n_x\times n_y)$ and the upsampling factor is $u$.

\subsection{Resolution}
The model image in \imshape is made at a resolution several times higher than the data resolution, and the convolution with the PSF also happens at this high resolution.  Only then is the model downsampled to compare to the data.  This procedure ensures that small and sharp features in the galaxy can be captured.

We find, though, that the central regions of galaxy bulges with the standard de Vaucouleurs profile are so sharp that an even higher resolution is needed to accurately model them.  Rather than generate the whole image at this extreme resolution we only use it for the central pixels where it is relevant.

There are therefore three different resolution parameters to consider: the upsampling (the ratio of the model to image pixel size in the outer image region), the size of the central higher resolution region, and the resolution increase in this central region (compared to the outer region).  The values that we used for these parameters are discussed in more detail in section \ref{upsampling-section}.

\section{Numerical optimization}
\label{sec:maximization}
The optimization problem of finding the best model parameters can be
tackled with a choice of algorithms; we have found that the best
performing is the Levenberg-Marquadt (LevMar) method.  This method
uses more information than methods (such as Powell's method or
Nelder-Meald) which operate only on the total likelihood as a function
of the input parameters: it uses the complete mismatch map, (i.e. the
$\chi^2$ per pixel) for each parameter set considered.

The LevMar method uses gradients (analytic or numerical) to optimize;
it is therefore not possible to signal a failure of a particular
parameter set (which can occur when parameters are near their limits)
using NaNs or other flag values.  While we do try to put in place a
parameterization that prevents this from happening (see below), we also
signal the failure by setting the $\chi^2$ for the failing model to 2
per pixel, a value large enough to be rejected but hopefully not
hugely problematic.

\subsection{Termination criteria}
\label{termination-criteria-section}
The LevMar method depends on a number of manually set thresholds that
determine when the numerical optimization should terminate. In
particular, it stops and reports the parameters of the best model fit
when at least one of the following criteria is met:
\begin{itemize}
\item One of the partial derivatives of the objective function
  ($\chi^2$) with respect to the model parameters falls below a threshold
  $\varepsilon_1$, i.e. $ \max | \frac{\partial \chi^2}{\partial p } |
  < \varepsilon_1 $, where $p$ denotes one of the model
  parameters. Note that this termination criterion can be sensitive to
  the changes of image pixel scaling and changes in the
  signal-to-noise ratio (SNR).
\item The relative change of the norm of the estimated parameter
  vector is smaller than $\epsilon_2$, i.e. $ \sum_p |p^{i-1}- p^i|^2 <
  \epsilon_2^2 \cdot \sum_p |p^{i-1}|^2$, where $i$ is the current iteration
  index. This termination criterion can be sensitive to individual
  parameter scaling.
\item The value of $\chi^2$ falls below $\epsilon_3$; again, this
  termination criterion is sensitive to image scaling and the S/N ratio.
\item The relative change of $\chi^2$ falls below $\epsilon_4$, i.e.
  $| (\chi^2)^{i-1} - (\chi^2)^{i} | / (\chi^2)^{i-1} < \epsilon_4/n$,
  where $(\chi^2)^{i}$ is the objective value at the current iteration
  and $n$ is the total number of pixels. This termination criterion
  was implemented by the authors and is not included in the LevMar
  implementation we are using. It can be
  sensitive to the step size that LevMar is taking and thus will be
  affected by the \texttt{levmar\_init\_mu} parameter - the initial
  damping of the normal equations solved in the Levenberg - Marquadt
  algorithm.
 \item The maximum number of iterations is reached (the minimization has failed).
\end{itemize}

\section{Analysis Overview}
\label{process-section}
Before analyzing a collection of galaxy images we:
\begin{enumerate}
	\item Read a collection of options specifying the model to be used, parameter ranges, and various options from a parameter file
	\item Build a model object encapsulating the model we will fit to the galaxies, including its likelihood and prior functions
\end{enumerate}

Before analyzing each individual image we then:
\begin{enumerate}
\item Generate the point-spread image based on PSF parameters; compute its FT.
\item Generate the sinc function for the specified upsampling (the FT of a top-hat).
\item Multiply these two Fourier space quantities together to generate the Fourier kernel.
\item Generate a weight map based on the image noise levels.
\item Compute a starting estimate for the galaxy properties from defaults and weighted quadrupole moments.
\end{enumerate}

We then pass the model likelihood function to a minimizer object which runs the chosen algorithm. At each evaluation of the posterior, for our standard model, we:
\begin{enumerate}
\item Check if the parameters are outside the desired ranges and give up early if not.
\item Build high-resolution (up-sampled) real-space models of each component of the galaxy (e.g. bulge and disc).
\item Replace the central few pixels of each model with summed up pixels from even higher resolution, since this is the region where the image changes quickly.
\item Sum and convolve the components with the precomputed kernel.
\item Down-sample the image to the data resolution.
\item Compute the likelihood from the weight map and model image.
\item Save the model image for use in the optimizer and residual analysis.
\end{enumerate}

\begin{figure}
\begin{center}
\includegraphics[width=60mm]{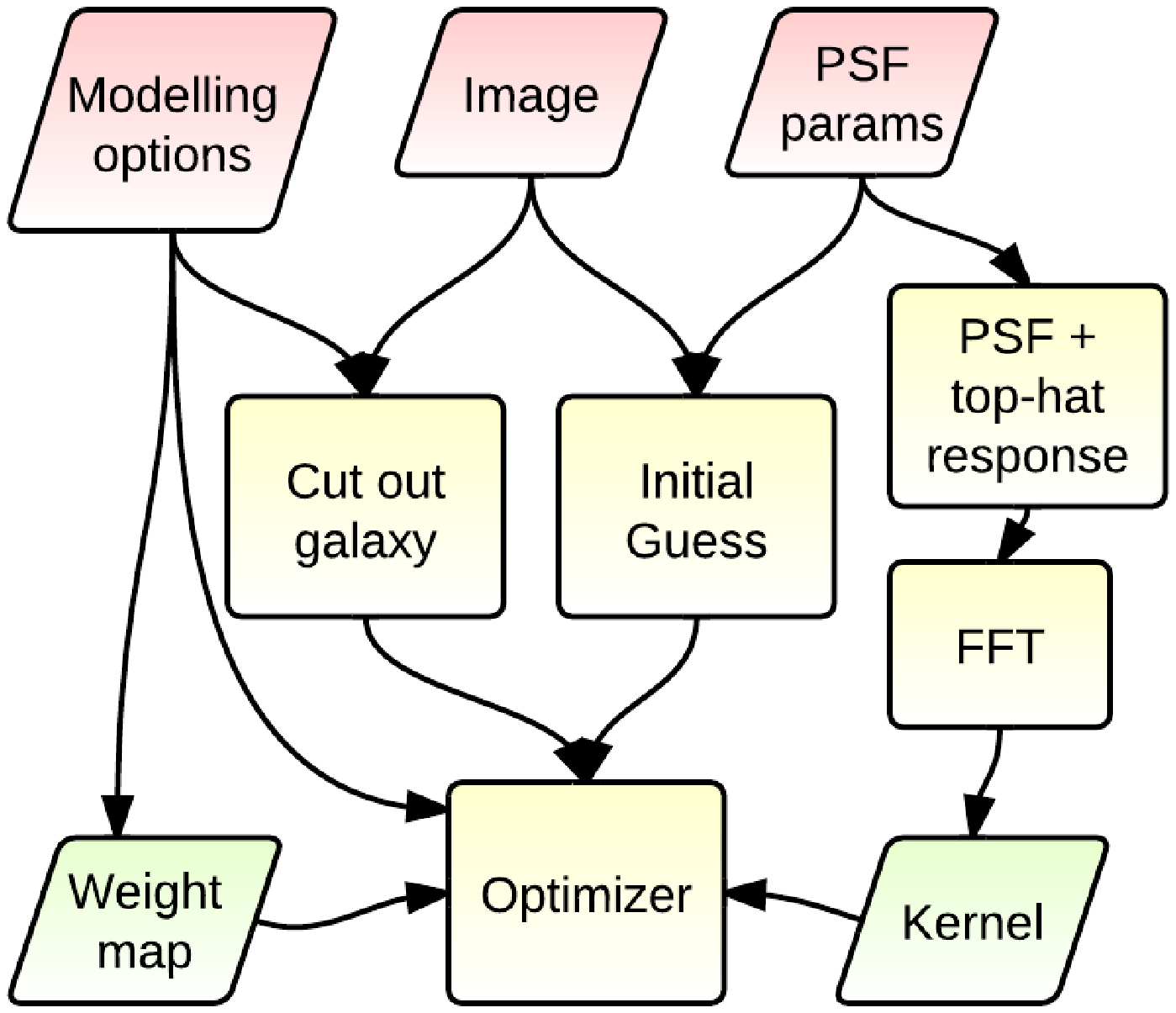}
\caption{
Actions in the \imshape code run to prepare for the analysis of a single galaxy.  Inputs, which can be collectively specified for a collection of galaxies, are upper red parallelograms, stored data are lower green parallelograms, and calculations are yellow rectangles.  The preparation time is negligible compared to the time of the optimization itself.}
\label{flow-chart-internal}
\end{center}
\end{figure}

\begin{figure}
\begin{center}
\includegraphics[width=60mm]{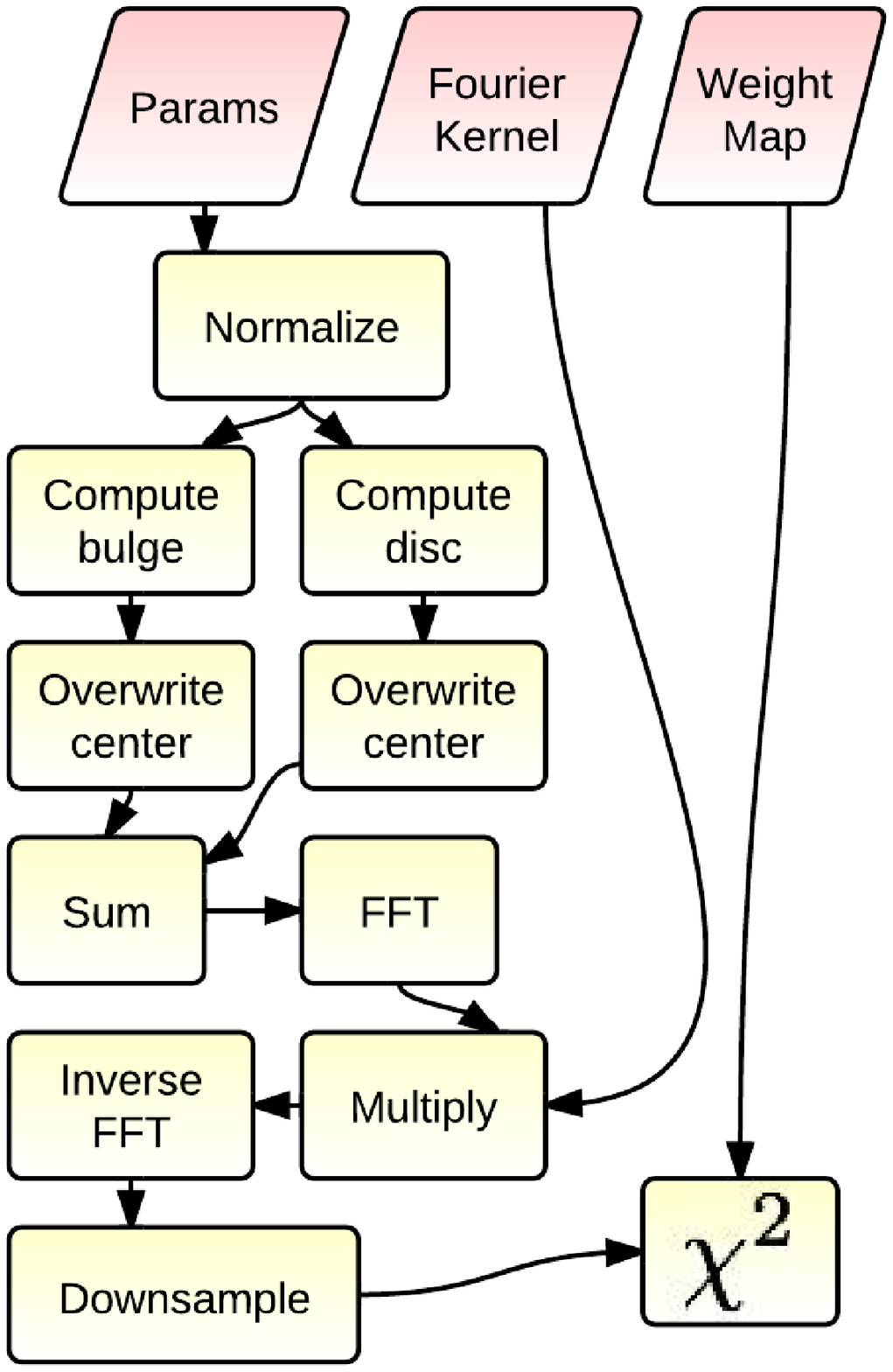}
\caption{The calculation performed whenever the optimization code provides a set of parameters for which to calculate the image and likelihood.  The principal parameterization used in this work and the normalization applied to it are described in section \ref{parameterization-section}.  The ``overwrite center'' process is described in more detail in section \ref{upsampling-section}.}
\label{flow-chart-usage}
\end{center}
\end{figure}

\newcommand{\data}{\mathbf{y}}
\newcommand{\psf}{\mathbf{f}}
\newcommand{\model}{\mathbf{x}}
\newcommand{\res}{\mathbf{r}}
\newcommand{\p}{\mathbf{p}}
\newcommand{\noise}{\mathbf{n}}
\newcommand{\Jac}{\mathbf{J}}
\newcommand{\cent}{\mathbf{d}}
\newcommand{\C}{\mathbf{C}}
\newcommand{\dist}{\mathbf{R}}

\section{Details on the maximum likelihood estimation}
\label{sec:jacobian}
The maximization problem outlined in section \ref{sec:maximization} is
based on the following model:
\begin{align}
\data = \mathrm{D} (\mathbf{K} \ast \model(\p)) + \noise, 
\label{eq:model}
\end{align}
where $\data$ denotes the observed galaxy image, $\mathbf{K}=\psf \ast \mathbf{\Pi}$, the convolution of the PSF $\psf$ and a square top-hat pixel window function $\mathbf{\Pi}$  (see section \ref{sec:top-hat}), $\model(\p)$ the (higher resolution) galaxy model with parameter
vector $\p$, and $\mathrm{D}(...)$ the downsampling operator, which averages into lower resolution pixels. As detailed
in section \ref{code section} we assume a two-component S\'{e}rsic
galaxy model with the parameters $\p=(x_0,y_0,e_1,e_2,r_d,A_b,A_d)$ (see
Table \ref{default parameter table} for a detailed parameter
description)
\begin{align}
  \model(\p)=\sum_{k=b,d} A_k \, \exp \left\{-\dist(x_0,y_0,e_1,e_2,r_d)^{1/2n_k}\right\},
\end{align}
where $n_k=1$ for the disc component and $4$ for the bulges, and $\dist$ denotes the sheared squared distance 
\begin{align}
 R_{i}=\cent_{i}^\intercal \mathrm{C} \,\cent_{i}
\label{eq:dist}
\end{align}
with the centered coordinate vector $\cent_{i}=(x_i-x_0,y_i-y_0)$ and the
covariance matrix 
\begin{align}
\mathrm{C}=\frac{1}{r_d (1-|e|)}\left(\begin{array}{cc} 1+|e|-2e_1 & -2e_2 \\ -2e_2 &
    1+|e|+2e_1\end{array}\right).
\label{eq:cov}
\end{align}
For additive Gaussian noise, i.e. $\noise \propto
\mathcal{N}(\mathbf{0},\mathbf{\Sigma})$ maximizing the likelihood
$P(\p|\data,\mathbf{K},\model)$ is equivalent to minimizing
\begin{align}
\chi^2(\p) = \frac{1}{2}  \sum_i r_i(\p)^2
\label{eq:optproblem}
\end{align}
with respect to the model parameters $\p$, where $\res$ denotes the
residual image between the observed galaxy image $\data$ and the
downsampled, PSF convolved galaxy model image $\model(\p)$, i.e.
\begin{align}
\res(\p) = \frac{1}{\mathbf{\sigma}}\,[\data - \mathrm{D}(\mathbf{K} \ast \model(\p))].
\end{align}
For simplicity we assume $\noise$ to be uncorrelated white noise and so consider only diagonal terms of
the covariance matrix, i.e. $\mathbf{\sigma} = \diag
\mathbf{\Sigma}$; we neglect any correlation in noise of
adjacent pixels.
 
For minimizing (\ref{eq:optproblem}), \imshape{} uses the free
software package {\sc levmar}\footnote{Available at
  \url{http://www.ics.forth.gr/~lourakis/levmar/}}, a C/C++
implementation of the Levenberg-Marquardt (LM) algorithm, a standard
technique for solving non-linear least-squares problems
\citep{presstvf92}. The LM method is based on a linear approximation
of $\res$ in the neighbourhood of $\p$, i.e.
\begin{align}
\res(\p + \Delta \p) \approx \res(\p) + \Jac(\p) \Delta \p +
O(\|\Delta \p\|^2), 
\end{align}
where $\Delta \p$ is a small perturbation of the parameter vector $\p$
and $\Jac$ denotes the Jacobian matrix of $\res$:
\begin{align}
\mathrm{J}_{ij}(\p) = \left.\frac{\partial r_i}{\partial p_j}\right|_\mathbf{p}.
\label{eq:jacmatrix}
\end{align}
The Jacobian (\ref{eq:jacmatrix}) can be numerically approximated, but to speed convergence of the minimization procedure it is
generally recommended to compute it analytically.  For our model
(\ref{eq:model}) the Jacobian reads with respect to the model
parameters $p \in \{x_0,y_0,e_1,e_2,r_d\}$
\begin{align}
\frac{\partial \res}{\partial p}  &=
\frac{1}{\mathbf{\sigma}}\mathrm{D}\left(\mathbf{K} \ast \sum_{k=b,d} \left[\frac{A_k}{2n_k}\,\exp  \left\{-\dist^{1/2n_k}\right\} \cdot \dist^{1/2n_k-1}
  \cdot \frac{\partial \dist}{\partial
    p}\right]\right)
\end{align}
where $\cdot$ denotes element-wise multiplication and the partial
derivatives $\partial \dist/\partial p$ are readily computed from
(\ref{eq:dist}) and (\ref{eq:cov}). With respect to the bulge and disc
amplitudes $A_{k\in\{b,d\}}$ one obtains
\begin{align}
\frac{\partial \res}{\partial A_k}  &=-\frac{1}{\mathbf{\sigma}}\,\mathrm{D}\left(\mathbf{K} \ast \exp \left\{-\dist^{1/2n_k}\right\}\right).
\end{align}

Note that for clarity and the ease of exposition we omitted the image
upsampling within a central image region as described in section
\ref{upsampling-section} in the derivation above, i.e.  we considered
the special case where ${\it n\_central\_pixel\_upsampling}$ is set to
one. The generalization to ${\it n\_central\_pixel\_upsampling}>1$ is
straightforward since it only involves an additional summation over
all ${\it n\_central\_pixel\_upsampling}^2$ sub-pixels to yield the
flux $[\model(\p)]_{ij}$ for each pixel $(i,j)$ within the central
image region.

\begin{figure}
\includegraphics[width=1.\columnwidth]{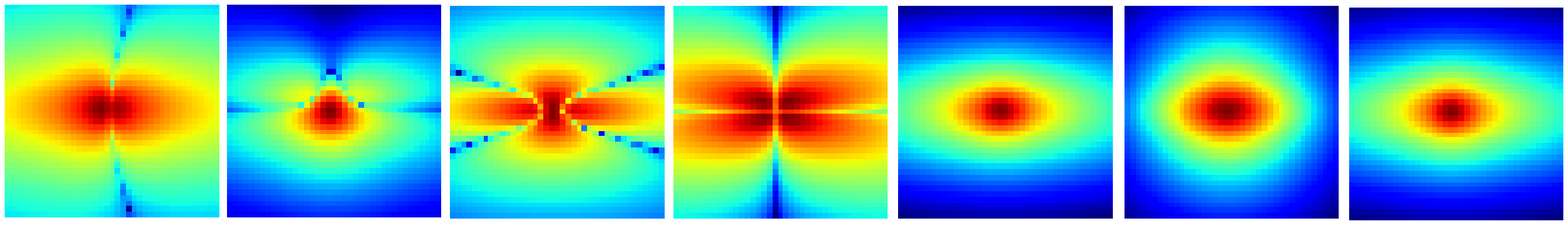}
\label{fig:jac}
\caption{Components of Jacobian matrix with respect to its model
  parameters (from left to right): $x_0$, $y_0$, $e_1$, $e_2$, $r_d$, $A_b$, $A_d$.}
\end{figure}

\section{Noise bias calibration using deep data}
\label{calibration appendix}
Noise bias is the difference between the expectation of the
maximum-likelihood result found by a model fitter like \imshape{},
which is a biased estimator, and the true underlying value, coming
from the non-linear mapping between parameters and image.  Its typical
value is a few percent, and we need to remove it to reach our target accuracy levels.

The size of the bias depends sensitively on the true parameters of the
galaxy, and if these were known we could remove noise bias completely.
However, we have access only to noisy estimates of these parameters
and therefore our estimate of the noise bias is itself noisy, and biased.  We can think of this as a \emph{bias-on-bias} problem, and we find that failing to account for it means we significantly miss our targets.

We can get around this problem if there is a subset of our
observational data that is deeper than the bulk of our sample, and
therefore of greater signal-to-noise, to a degree sufficient that it
has negligible noise bias.  This is often the case in
real surveys that seek to detect high redshift supernovae, and it is approximated in the GREAT08 challenge with LowNoise\_Blind data set (although the galaxy types in this set do not match those in the main sample exactly).  In this case we do not calibrate each galaxy individually, but instead find a mean bias for the population and apply it en masse.

Biases $m(\theta)$ and $c(\theta)$ will afflict each of our galaxies, depending on the true galaxy properties $\theta$.  We can calculate mean values of this bias $\hat{m}$ and $\hat{c}$, and apply these evenly to all the galaxies, such that the mean galaxy ellipticity (which is our goal) is correct.

The mean of the bias across the population is given by:
\begin{equation}
\hat{m} = \int m(\theta) p(\theta) \mathrm{d}\theta.
\end{equation}
We find the distribution $p(\theta)$ using fits to the deep data, and $m(\theta)$ using simulations - for each point $\theta_p$ in a grid in the parameter space  we simulate many galaxies and determine the value $m(\theta_p)$.  We can then interpolate between these values to do the integral, and finally apply the mean $\hat{m}$ to all the galaxy estimates.  

A similar process is used for the $c$ bias, except that $c$ varies directly with the PSF, so we use this information - we calculate the $\hat{c}$ assuming a fiducial PSF with ellipticity $e^\mathrm{psf}_0$ aligned with the $e_1$ direction. The applied value to apply to each galaxy is then:
\begin{equation}
c_1 + i c_2 = \hat{c} \cdot \frac{e^\mathrm{psf}  }{e^\mathrm{psf}_0} \cdot  e^{i \theta_\mathrm{psf}}.
\end{equation}

\section{Results tables}
Tables \ref{Q table}-\ref{c2 table} give the GREAT08 $Q$, $m$, $c_1$, and $c_2$ values for \imshape{} with (I3) and without (I3-U) noise bias calibration applied.  Also shown are the scores for the other entrants at the time of the challenge, as described in \citet{great08results}.

\begin{table}
\center
\caption{Quality factor $Q$ scores for GREAT08 from this work and the original challenge entrants.}
\label{Q table}
\begin{tabular}{| c | c c c c c c c c c|}
\hline
 & \multicolumn{9}{c}{Branch} \\
Name & 1 & 2 & 3 & 4 & 5 & 6 & 7 & 8 & 9 \\
\hline
      I3 & 817.2 & 1342.7 & 450.8 & 725.1 & 722.0 & 4702.5 &  84.6 &  382.7 &  99.8  \\
    I3-U & 128.7 &  183.5 &  67.2 &  78.1 & 115.4 & 1417.1 &  25.3 & 2613.0 &  10.6  \\
      HB & 256.4 &  654.4 & 241.9 & 184.2 & 158.9 &  225.3 & 176.0 &  272.1 & 131.7  \\
      AL & 764.9 &  595.1 & 348.4 & 121.9 & 404.0 &  282.8 &  22.4 &  529.5 & 512.1  \\
      TK & 224.7 &  569.6 & 737.4 & 161.9 & 238.5 &   63.6 &  54.3 &  142.9 &  59.9  \\
      CH & 174.2 &  324.7 & 157.4 &  83.3 & 212.4 &  173.9 &  13.1 &  230.8 &  18.6  \\
      PG &  21.7 &   26.5 &  26.3 &  37.8 &  44.2 &   28.5 &  50.6 &   23.1 &  82.4  \\
      MV & 246.2 &  155.3 &  95.1 & 126.3 & 224.4 &  488.7 &   4.3 &  398.8 &  23.8  \\
      KK & 222.2 &  554.8 & 448.6 &  89.8 & 230.4 &  100.0 &   3.1 &  163.4 &  33.7  \\
    HHS3 &  26.9 &   33.0 &  30.7 &  10.2 &  20.3 &   30.2 &  30.3 &   20.5 &  25.2  \\
      SB & 161.4 &   93.4 & 226.3 & 114.5 & 155.2 &   19.8 &   4.9 &   26.0 &   8.6  \\
    HHS2 &  23.0 &   26.8 &  26.1 &   9.9 &  18.3 &   23.4 &  25.4 &   19.4 &  23.2  \\
    HHS1 &  67.2 &  109.0 &  22.3 &  43.0 &  60.9 &  232.7 &   8.8 & 1516.6 &   2.1  \\
      MJ &  10.6 &   11.5 &  10.4 &  23.0 &   4.0 &    8.3 &  13.9 &   10.4 &  16.4  \\
    USQM &  13.3 &   15.8 &  11.4 &   9.4 &  23.4 &    8.4 &   0.8 &    1.1 &   0.2  \\
\hline
\end{tabular}
\end{table}

\begin{table*}
\center
\caption{Multiplicative bias $m$ scaled by $10^{-3}$ on GREAT08 from this work and the original challenge entrants.}
\label{m table}
\begin{tabular}{| c | c c c c c c c c c|}
\hline
 & \multicolumn{9}{c}{Branch} \\
Name & 1 & 2 & 3 & 4 & 5 & 6 & 7 & 8 & 9 \\
\hline
      I3 &  15.3 &   3.3 &  -3.3 &  12.9 &  13.9 &    3.4 &  -35.8 &  -18.3 &  91.9 \\
    I3-U &  37.6 &  25.4 &  49.8 &  35.1 &  36.2 &    8.9 &   49.0 &    3.3 & 116.0 \\
      HB &  28.9 &  12.8 &  25.6 &  23.3 &  24.9 &   28.0 &   11.3 &   25.6 &  37.8 \\
      AL &   9.9 &  -3.1 &   7.8 &   4.9 &   6.5 &   23.3 &  -87.5 &   -3.4 &   8.4 \\
      TK &  26.6 &  14.3 &  -5.6 &  20.9 &  21.1 &   59.6 &  -57.0 &   37.0 &  -3.4 \\
      CH &  17.8 &   7.9 &  11.0 &  11.7 &  16.7 &    9.5 &  -35.1 &   24.6 &   0.1 \\
      PG &  70.3 &  58.1 &  58.9 &  70.5 &  67.0 &   38.3 &  -38.6 &   61.4 &  42.1 \\
      MV & -27.4 & -35.8 & -43.2 & -36.6 & -30.0 &   19.6 & -188.4 &  -19.3 & -84.4 \\
      KK &  25.7 &   8.8 &   5.0 &  16.8 &  21.6 &   45.4 & -254.4 &   27.6 & -61.0 \\
    HHS3 &  74.1 &  64.1 &  67.0 &  70.2 &  69.8 &   66.2 &   63.8 &   87.6 &  74.5 \\
      SB & -36.9 & -48.1 & -22.6 & -41.6 & -35.6 & -101.2 & -167.4 &  -88.4 &  -7.3 \\
    HHS2 &  83.1 &  70.7 &  72.7 &  75.5 &  79.2 &   79.1 &   75.8 &   90.4 &  82.5 \\
    HHS1 &  54.5 &  39.1 &  89.6 &  43.7 &  46.6 &  -28.1 &  150.7 &   -8.9 & 306.0 \\
      MJ &  61.1 &  47.1 &  58.0 &  54.9 &  53.3 &  -36.5 &   22.5 &    3.1 &  26.6 \\
    USQM & -52.5 & -60.0 & -62.7 & -51.3 & -49.2 & -118.6 &  420.8 & -393.5 & 882.2 \\
\hline
\end{tabular}
\end{table*}

\begin{table*}
\center
\caption{Additive bias $c_1$ scaled by $10^{-5}$ on GREAT08 from this work and the original challenge entrants.}
\label{c1 table}
\begin{tabular}{| c | c c c c c c c c c|}
\hline
 & \multicolumn{9}{c}{Branch} \\
Name & 1 & 2 & 3 & 4 & 5 & 6 & 7 & 8 & 9 \\
\hline
      I3 &  -22.0 &  -31.8 &   63.8 &  -33.9 &   12.8 &  -17.8 &  -52.5 &   44.6 & -254.3 \\
    I3-U &  -67.0 &  -77.0 &  -94.6 & -124.0 &   29.4 &  -29.0 & -235.0 &    1.1 & -304.0 \\
      HB &   -4.4 &  -17.6 &  -16.8 &  -66.3 &   58.2 &  -45.2 &   59.4 &  -26.1 &   10.7 \\
      AL &  -38.4 &  -51.5 &  -59.3 & -118.6 &   34.8 &  -54.5 &  -79.9 &  -53.3 &  -51.3 \\
      TK &  -26.3 &  -28.1 &   20.5 &  -63.5 &    9.5 &  -50.7 &   13.4 &  -44.4 &  154.6 \\
      CH &   82.3 &   61.7 &   84.3 &  143.2 &  -22.0 &   93.1 &   -9.9 &   23.7 &  262.5 \\
      PG &  188.4 &  177.9 &  182.1 &  -41.9 &  -87.5 &  220.0 &  -66.8 &  186.5 &   34.1 \\
      MV &    8.5 &   -0.9 &  -37.5 &  -31.6 &   23.1 &   -6.7 & -283.5 &   38.0 &  -79.3 \\
      KK &  -49.2 &  -44.6 &  -47.1 & -132.2 &   37.6 &  -57.8 & -119.5 &  -62.8 & -129.9 \\
    HHS3 & -174.2 & -165.6 & -160.1 & -387.2 &   75.8 & -180.0 & -168.8 & -191.8 & -196.5 \\
      SB &    9.5 &   16.6 &  -54.5 &   18.7 &   16.0 &  135.4 & -311.2 &  110.3 & -459.1 \\
    HHS2 & -177.6 & -185.8 & -176.0 & -385.1 &   86.0 & -196.5 & -178.0 & -195.3 & -189.0 \\
    HHS1 &  -71.2 &  -74.8 & -138.6 & -169.4 &   56.4 &   48.4 & -225.2 &    1.2 & -431.7 \\
      MJ & -245.8 & -249.7 & -219.1 &   69.1 & -696.9 & -345.8 &  -92.1 & -288.8 &  285.7 \\
    USQM & -138.0 & -111.2 & -114.0 & -295.7 &   70.3 & -118.8 & -149.5 & -161.5 &  -20.2 \\
\hline
\end{tabular}
\end{table*}

\begin{table*}
\center
\caption{Additive bias $c_2$ scaled by $10^{-5}$ on GREAT08 from this work and the original challenge entrants.}
\label{c2 table}
\begin{tabular}{| c | c c c c c c c c c|}
\hline
 & \multicolumn{9}{c}{Branch} \\
Name & 1 & 2 & 3 & 4 & 5 & 6 & 7 & 8 & 9 \\
\hline
      I3 &    2.1 & -10.6 &  -17.0 &   9.1 &   3.4 &  -0.6 &    9.4 & -24.4 &    61.2 \\
    I3-U &   18.5 &   5.6 &   41.7 &  42.1 &  47.9 &   3.5 &   75.8 &  -8.5 &    78.9 \\
      HB &   15.3 &  11.6 &   31.2 &  43.2 &  28.1 &  19.0 &   31.6 &  15.3 &    -1.6 \\
      AL &   19.6 &  11.0 &   39.7 &  49.6 &  49.3 &  15.8 &  -14.1 &  10.7 &    -1.0 \\
      TK &   44.6 &  30.8 &   40.0 &  69.8 &  61.5 &  12.5 &   28.1 &  24.5 &   -92.7 \\
      CH &  -16.2 & -37.5 &    6.1 & -39.4 & -72.4 & -33.0 &   53.4 & -37.7 &   -24.0 \\
      PG &   21.2 &  14.7 &   39.0 &  81.9 &  -9.0 &  11.0 &   79.9 &   5.0 &    40.3 \\
      MV &   14.4 & -14.0 &   26.9 &  27.7 &  -4.5 & -11.1 &   82.0 &  -1.5 &     1.9 \\
      KK &   18.7 &   1.7 &   41.8 &  60.1 &  39.4 &   9.6 &   10.2 &  19.6 &   -24.4 \\
    HHS3 &   54.9 &  63.1 &   83.7 & 148.3 & 193.6 &  57.3 &   70.9 &  69.6 &    56.5 \\
      SB &  -15.1 & -27.0 &   22.3 & -32.8 & -45.2 & -53.7 &   87.5 & -43.4 &   126.8 \\
    HHS2 &   69.4 &  65.4 &   84.5 & 150.6 & 185.9 &  67.3 &   58.1 &  69.3 &    48.3 \\
    HHS1 &   27.1 &  18.2 &   79.0 &  72.4 &  77.0 & -20.1 &   49.2 &  -4.7 &   145.2 \\
      MJ &  322.9 & 316.9 &  349.5 & 218.3 & 127.9 & 313.0 &  353.2 & 329.2 &   164.7 \\
    USQM & -114.4 & -54.5 & -182.7 & -83.3 & -33.3 &  22.0 & -977.9 & 187.4 & -1615.3 \\
\hline
\end{tabular}
\end{table*}

\end{document}